# Inapproximability of Dominating Set in Power Law Graphs

Mikael Gast* Mathias Hauptmann† Marek Karpinski‡

**Abstract**

We give logarithmic lower bounds for the approximability of the MINIMUM DOMINATING SET problem in connected $(\alpha, \beta)$-Power Law Graphs. We give also a best up to now upper approximation bound on the problem for the case of the parameters $\beta > 2$. We develop also a new functional method for proving lower approximation bounds and display a sharp phase transition between approximability and inapproximability of the underlying problem. This method could also be of independent interest.

**Keywords:** Approximation Algorithms, Inapproximability, Power Law Graphs, Combinatorial Optimization, Dominating Set

## 1 Introduction

Recent developments in the analysis of large scale real-world networks often reveal common topological signatures and statistical features that are not easily captured by classical *uniform* random graphs—such as generated by the $G(n, p)$-model due to Erd??s and Rényi [ER60]. One of the crucial observations is that the distribution of node degrees is well approximated by a *power law distribution*, i.e. the number of nodes $y_i$ of a given degree $i$ is proportional to $i^{-\beta}$ where $\beta > 0$. This was verified experimentally for a large number of existing real-world networks such as the Internet, the World-Wide Web, protein-protein interaction networks, gene regulatory networks, peer-to-peer networks, mobile call networks, et cetera [FFF99; Kle+99; Kum+00; Bro+00; KL01; JAB01; Gue+02; Sig+03; Eub+04b; Ses+08].

In the research of epidemic spreading of diseases across networks of travel routes and networks of social contacts [PV01; Eub+04a] or the broadcasting of information inside large wireless networks [SSZ02], a natural question arises about how to efficiently place key nodes at key positions inside a network such as to reach and to effect all or most of the remaining nodes. Here, the feasibility of a solution also heavily depends on the number of key nodes needed in order to cover the whole network and thus this number is often tried to be minimized. Questions like these quickly resemble or are equivalent to classical NP-hard optimization problems, i.e. minimum covering and domination problems in the context of graph theory.

In connection with the optimal placement of sensors for disease detection inside social networks Eubank et al. [Eub+04b] studied near-optimal MINIMUM DOMINATING SET problems ($(1-\varepsilon)$-MIN-DS) in *bipartite random power law graphs.* On a graph $G = (V, E)$ a *dominating set* is a subset $D \subseteq V$ such that every node in $V \setminus D$ is connected to $D$ by some edge in $E$. MIN-DS asks for a dominating set of minimum cardinality $|D|$. This problem is known to be

*Dept. of Computer Science, University of Bonn and B-IT Research School. e-mail: gast@cs.uni-bonn.de
†Dept. of Computer Science, University of Bonn. e-mail: hauptman@cs.uni-bonn.de
‡Dept. of Computer Science and the Hausdorff Center for Mathematics, University of Bonn. Research partially supported by the Hausdorff grant EXC59-1. e-mail: marek@cs.uni-bonn.de

NP-hard by a reduction from the Set Cover problem and the result of Raz and Safra [RS97] rules out the existence of an approximation algorithm for general graphs with an approximation factor better than $c \cdot \log |V|$ for some $c > 0$ unless P = NP. Eubank et al. found that for a class of bipartite random power law graphs the problem $(1-\varepsilon)$-Min-DS is easier to solve, i.e. they presented a simple greedy algorithm which achieves a $(1 + o(1))$-approximation on these instances.

Ferrante, Pandurangan, and Park [FPP08] and Shen, Nguyen, and Thai [SNT10; She+12] studied the approximation hardness of Minimum Vertex Cover (Min-VC), Maximum Independent Set (Max-IS) and Minimum Dominating Set (Min-DS) in *combinatorial power law graphs* and showed NP-hardness and APX-hardness for simple $(\alpha, \beta)$-PLG and $(\alpha, \beta)$-PLG multigraphs, respectively. In Table 1 we list some of these results, especially the previously best lower bound for Minimum Dominating Set in $(\alpha, \beta)$-PLG for $\beta > 0$.

| Problem | $(\alpha, \beta)$-PLG multigraphs | $(\alpha, \beta)$-PLG |
|---|---|---|
| Max-IS | $1 + \frac{1}{140(2\zeta(\beta)3^\beta - 1)} - \varepsilon$ | $1 + \frac{1}{1120\zeta(\beta)3^\beta} - \varepsilon$ |
| Min-DS | $1 + \frac{1}{390(2\zeta(\beta)3^\beta - 1)}$ | $1 + \frac{1}{3120\zeta(\beta)3^\beta}$ |
| Min-VC | $1 + \frac{2\left(1-(2+o_c(1))\frac{\log\log c}{\log c}\right)}{\left(\zeta(\beta)c^\beta + c^{\frac{1}{\beta}}\right)(c-1)}$ | $1 + \frac{2-(2+o_c(1))\frac{\log\log c}{\log c}}{2\zeta(\beta)c^\beta(c+1)}$ |

**Table 1:** Previously known lower bounds for the approximability of Max-IS and Min-DS in PLG under condition P $\neq$ NP, Min-VC under UGC in disconnected power law graphs with $\beta > 1$ due to Shen et al. [She+12].

The underlying combinatorial model for power law graphs was proposed by Aiello, Chung, and Lu in [ACL00; ACL01], we refer also a general reader to this reference. According to their definition, an $(\alpha, \beta)$-power law graph is an undirected (multi-)graph with maximum degree $\Delta = \left\lfloor e^{\alpha/\beta} \right\rfloor$, which contains for each $1 \leqslant i \leqslant \Delta$ $y_i$ nodes of degree $i$, where

$$y_i = \begin{cases} \left\lfloor \frac{e^\alpha}{i^\beta} \right\rfloor & \text{if } i > 1 \text{ or } \sum_{i=1}^{\Delta} \left\lfloor \frac{e^\alpha}{i^\beta} \right\rfloor \text{ is even} \\ \lfloor e^\alpha \rfloor + 1 & \text{otherwise.} \end{cases}$$

Here, $i$ and $y_i$ satisfy $\log y_i = \alpha - \beta \log i$. Furthermore, $\alpha$ is the logarithm of the size of the graph and $\beta$ is the log-log growth rate. In [ACL00; ACL01], also a random graph model for $(\alpha, \beta)$-PLG is proposed which is based on a random matching construction. We refer to this random graph model as the *ACL-model*.

## 2 Summary of Our Results

In this paper we study the approximation complexity of Min-DS in $(\alpha, \beta)$-PLG. We give logarithmic lower bounds for the approximability of this problem for $0 < \beta \leqslant 2$, based on a reduction from the Set Cover problem combined with the logarithmic lower bound for Set Cover given by U. Feige [Fei98]. The previously known results were the constant factor lower bounds given in [She+12], which were based on reductions from the bounded degree Min-DS. It was also shown in [She+12] that for $\beta > 2$, Min-DS in $(\alpha, \beta)$-PLG is in APX. We improve on this result by giving new upper bounds on the approximation ratio of an algorithm based on the greedy algorithm for Min-DS. In [She+12], membership of Min-DS in $(\alpha, \beta)$-PLG in APX was shown by constructing a lower bound for the optimum and an upper bound for the greedy solution separately. We obtain our new results by relating the cost and structure of an optimum solution to those of a greedy-based solution. This sophisticated analysis yields improved upper bounds for almost the whole range $\beta > 2$. Finally we take a very close look at the phase transition at $\beta = 2$. Similar as in [GHK12] we extend the power law model and consider the case when $\beta_f = 2 + \frac{1}{f(n)}$ is a function of the graph size $n$ which converges to 2 from above. We obtain the following surprising result: For every function $f(n)$ with $f(n) = \omega(\log(n))$ (i.e. when $\beta_f$ converges fast enough), MIN-DS in $(\alpha, \beta_f)$-PLG still provides a logarithmic approximation lower bound, and for every function $f(n)$ with $f(n) = o(\log(n))$, the problem is in APX. The summary of main results of this paper is given in Table 2.

| | Approximation Lower Bound |
|---|---|
| $0 < \beta < 1$ | $\Theta\left(\ln(n) - \ln\left(\frac{1}{1-\beta}\right)\right)$ |
| $\beta = 1$ | $\Theta\left(\ln(n)\right)$ |
| $1 < \beta < 2$ | $\Theta\left(\ln(n) - \ln(\zeta(\beta))\right)$ |
| $\beta = 2$ | $\Theta\left(\ln(n) - \ln(\zeta(\beta))\right)$ |
| $\beta = 2 + \frac{1}{f(n)}, f(n) = \omega(n)$ | $\Theta\left(\ln(n) - \ln(\zeta(\beta))\right)$ |

| | Approximation Upper Bounds |
|---|---|
| $\beta = 2 + \frac{1}{f(n)}, f(n) = o(n)$ | APX |
| $\beta > 2$ | APX |
| $\beta > 2.729$ | $\frac{\zeta(\beta) - \frac{\zeta(\beta-1)}{2}}{1 - \frac{\zeta(\beta-1)}{2}}$ |

**Table 2:** Summary of our main results.

# 3 Organization of the Paper

In section 4 we give an outline of our methods and the embedding constructions on which our reductions are based. In section 5 we take a close look at U. Feige's original reduction from 5OCC-MAX-E3SAT (5 OCCURRENCE MAXIMUM E3SAT) to the SET COVER [Fei98] and the standard reduction from the SET COVER to the MINIMUM DOMINATING SET problem. As a result of this section, we obtain sufficient information about the degree distribution of the resulting MIN-DS instances $G_{U,\mathcal{S}}$. In section 6 we give new lower bounds on the approximability of MIN-DS in $(\alpha, \beta)$-PLG. The subsection 6.1 deals with the case $1 < \beta < 2$. We describe how to *rescale* the degree distribution of instances $G_{U,\mathcal{S}}$ in order to embed them into an $(\alpha, \beta)$-PLG. We also apply our scaling technique for the case $\beta = 2$ in subsection 6.4 together with a slightly different analysis. The case $0 < \beta < 1$ is treated in subsection 6.2, based on a precise rounding error analysis for the terms that determine the lower approximation bound. An similar analysis is used for the case $\beta = 1$ in subsection 6.3.

In section 7 we present new upper bounds for the case of $\beta > 2$ and provide a detailed comparison of the previous and new upper bounds in terms of the parameter $\beta$.

In section 8 we consider the functional case when $\beta_f = 2 + \frac{1}{f(n)}$ is a function of the graph size $n$ which converges to 2 from above.

## 4 Outline of the Method

Let us give an outline of our methods and constructions.

In order to obtain logarithmic approximation lower bounds for the Minimum Dominating Set problem in $(\alpha, \beta)$-*Power Law Graphs*, we construct reductions from Min-DS in graphs, which is basically as hard to approximate as the Set Cover problem. It is well known [PM81; BM84; Kan92] that Set Cover instances $U, \mathcal{S}$ with universe $U$ and set system $\mathcal{S}$ can be translated into instances $G_{U,\mathcal{S}}$ of Min-DS in graphs, where $G_{U,\mathcal{S}}$ contains a vertex for every element of $U$ and vertices for the sets $S \in \mathcal{S}$. Element vertices are connected to set vertices of those sets in which they are contained, and two set vertices are connected by an edge if and only if the two sets have a non-empty intersection.

Our reductions map those graphs $G_{U,\mathcal{S}}$ which are stemming from Set Cover instances $U, \mathcal{S}$ to $(\alpha, \beta)$-Power Law Graphs $\mathcal{G}_{\alpha,\beta}$. In this construction, nodes of the graph $G_{U,\mathcal{S}}$ are connected to a set $\Gamma$ of degree-2 nodes, and those are again connected to the rest of the graph. The set $\Gamma$ enforces any *reasonable* dominating set in $\mathcal{G}_{\alpha,\beta}$ to contain a dominating set of the graph $G_{U,\mathcal{S}}$.

Another important property of our constructions is that the residual graph $\mathcal{G}_{\alpha,\beta} \setminus (G_{U,\mathcal{S}} \cup \Gamma)$ contains a sufficiently small set $X$ of vertices which dominate every node in $\mathcal{G}_{\alpha,\beta} \setminus G_{U,\mathcal{S}}$. It is precisely this property which enables us to obtain logarithmic lower bounds (instead of the previously known constant lower bounds) for the approximability of Min-DS in $(\alpha, \beta)$-PLG.

The crucial point in this construction is the implementation of the *power law distribution*. Therefore we need to know the degree distribution in the graph $G_{U,\mathcal{S}}$. In section 5 we will take a close look at Feige's original construction and obtain upper and lower bounds for the degrees of nodes in the graph $G_{U,\mathcal{S}}$, where $(U, \mathcal{S})$ is a Set Cover instance in Feige's construction. We apply our construction only to those Set Cover instances $(U, \mathcal{S}) = F_{SC}(\varphi)$ where $\varphi$ is a 5OCC-MAX-E3SAT instance and $F_{SC}$ is Feige's reduction from [Fei98]. We show that the Min-DS instances $G_{U,\mathcal{S}}$ have the following property: There exist constants $0 < a < b < 1$ such that for every $(U, \mathcal{S})$ with $(U, \mathcal{S}) = F_{SC}(\varphi)$, the node degrees of all vertices in $G_{U,\mathcal{S}}$ are contained in the interval $\left[N^a, N^b\right]$, where $N$ is the number of vertices of $G_{U,\mathcal{S}}$.

**Intervals and Volumes.** In an $(\alpha, \beta)$-Power Law Graph $\mathcal{G}_{\alpha,\beta} = (V_{\alpha,\beta}, E_{\alpha,\beta})$ the number $|V_{\alpha,\beta}| = n = \sum_{i=1}^{\lfloor e^{\alpha/\beta} \rfloor} \left\lfloor \frac{e^\alpha}{i^\beta} \right\rfloor$ of nodes and the number $|E_{\alpha,\beta}| = m = \frac{1}{2} \sum_{i=1}^{\lfloor e^{\alpha/\beta} \rfloor} \left\lfloor \frac{e^\alpha}{i^\beta} \right\rfloor$ of edges satisfy

$$n \approx \begin{cases} \zeta(\beta) e^\alpha & \text{if } \beta > 1 \\ \alpha e^\alpha & \text{if } \beta = 1 \\ \frac{e^{\frac{\alpha}{\beta}}}{1-\beta} & \text{if } 0 < \beta < 1 \end{cases} \quad \text{and} \quad m \approx \begin{cases} \frac{1}{2}\zeta(\beta - 1) e^\alpha & \text{if } \beta > 2 \\ \frac{1}{4}\alpha e^\alpha & \text{if } \beta = 2 \\ \frac{1}{2}\frac{e^{\frac{2\alpha}{\beta}}}{2-\beta} & \text{if } 0 < \beta < 2 \end{cases}$$

In the following we will introduce notations of *intervals* of nodes inside an $(\alpha, \beta)$-Power Law Graph and of the *volume* of such an interval. Let $\mathcal{G}_{\alpha,\beta} = (V, E)$ be an $(\alpha, \beta)$-PLG. An interval of nodes in $\mathcal{G}_{\alpha,\beta}$ is a set $[a, b] = \{v \in V \mid a \leqslant \mathsf{deg}(v) \leqslant b\}$, where $1 \leqslant a \leqslant b \leqslant \Delta = \left\lfloor e^{\alpha/\beta} \right\rfloor$. Furthermore, let $|[a, b]|$ be the number of nodes inside the interval $[a, b]$. For the volume of an interval $[a, b]$ we define $\mathrm{vol}([a, b]) = \sum_{j=a}^{b} \left\lfloor \frac{e^\alpha}{j^\beta} \right\rfloor \cdot j$, i.e. sum of node degrees of nodes inside the interval.

**Embedding Technique.** Here we construct a map which embeds every graph $G_{U,\mathcal{S}}$ (where $U, \mathcal{S}$ is a Set Cover instance from Feige's hardness result) into an $(\alpha, \beta)$-PLG $\mathcal{G}_{\alpha,\beta}$. Let $G_{U,\mathcal{S}} = (V_{U,\mathcal{S}}, E_{U,\mathcal{S}})$ with $|V_{U,\mathcal{S}}| = N$. The graphs $G_{U,\mathcal{S}}$ have the following property: There exist constants $0 < a < b < 1$ such that for all $v \in V_{U,\mathcal{S}}$, $N^a \leqslant \mathsf{deg}_{U,\mathcal{S}}(v) \leqslant N^b$. The power law graph $\mathcal{G}_{\alpha,\beta} = (V_{\alpha,\beta}, E_{\alpha,\beta})$ has the vertex set $V_{\alpha,\beta} = V_{U,\mathcal{S}} \cup X \cup \Gamma \cup V_1 \cup W$, where $X \subseteq [x\Delta, y\Delta] = \{v \in V_{\alpha,\beta} | \, x\Delta \leqslant \mathsf{deg}_{\alpha,\beta}(v) \leqslant y\Delta\}$ is the set of dominating nodes, $V_1$ is the set of degree-1 nodes and $W$ the set of remaining nodes of the targeted degree sequence. $\mathcal{G}_{\alpha,\beta}$ is constructed such that each node in $V_{U,\mathcal{S}}$ has precisely one neighbor in $\Gamma \subseteq W$, and every $u \in \Gamma$ has precisely one neighbor in $V_{U,\mathcal{S}}$. Furthermore, each node $w \in W$ is adjacent to precisely one node in $X$ and every degree-1 node is adjacent to a node in $X$, where each $v \in X$ has at least one degree-1 neighbor. The set $X$ is chosen to dominate every vertex in $W$ and all the degree-1 nodes in $V_1$ (cf. Figure 1).

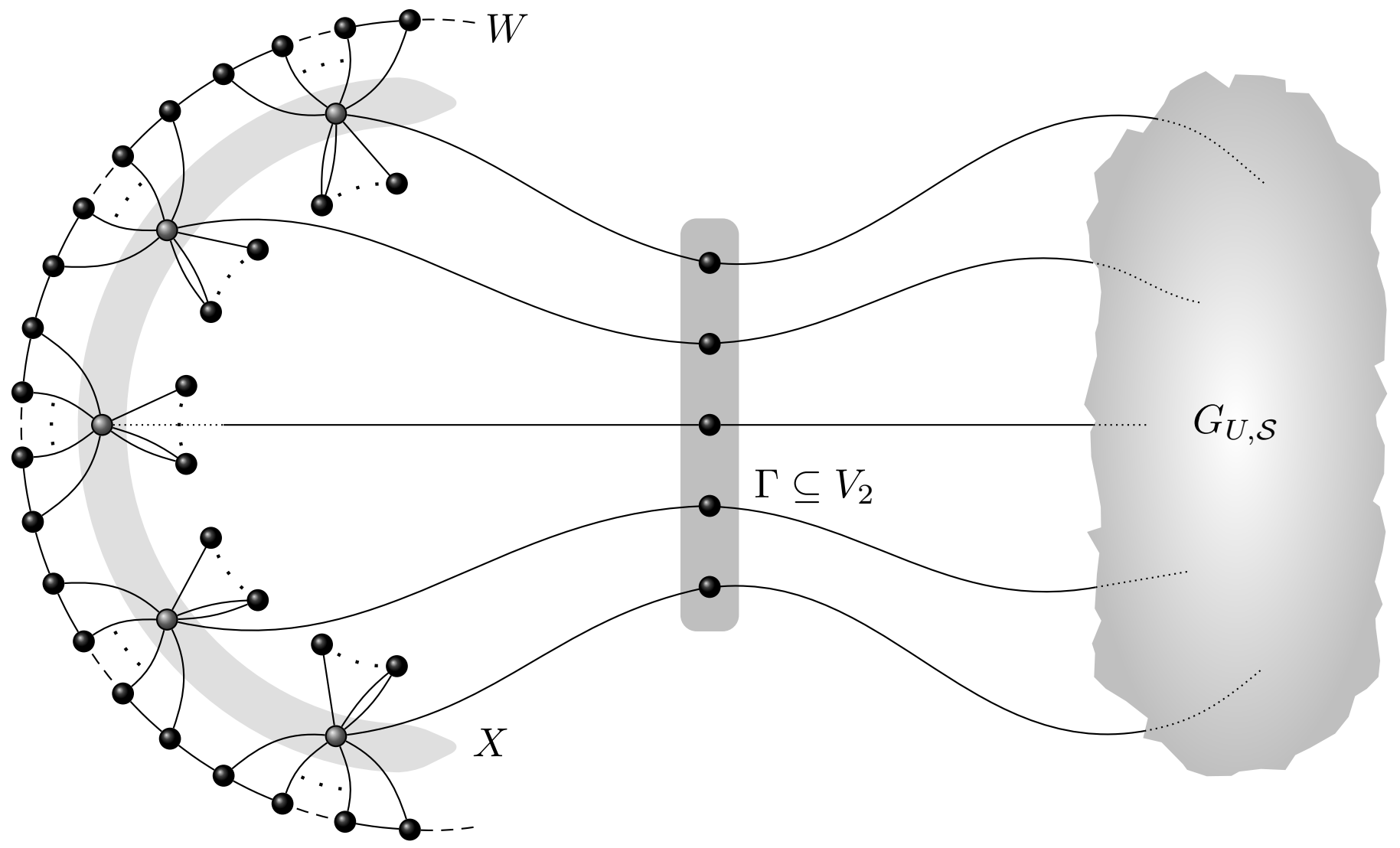

**Figure 1:** The main construction for the embedding of a MIN-DS (SET COVER) instance $G_{U,\mathcal{S}}$ into a $(\alpha, \beta)$-PLG. In the resulting graph the nodes $\bullet \in X$ are dominating the sets $W \cup V_1$, separating the dominating set in $G_{U,\mathcal{S}}$ from the dominating set in $\mathcal{G}_{\alpha,\beta} \setminus G_{U,\mathcal{S}}$.

In order to be able to monitor the current status of implementing the power-law degree distribution inside the graph $\mathcal{G}_{\alpha,\beta}$, we keep track of the *residual degrees* $\mathsf{deg}_r(v)$ of nodes in $X \cup W \cup V_1$. Consider the algorithm `ConstructPLG`.

The last two steps of the algorithm `ConstructPLG` are calling the procedure `Fill_Wheel` on the sets which may still have residual degrees (because of space constraints here we refer for the procedure `Fill_Wheel` to [GHK12, p. 8]). The procedure `Fill_Wheel` gets as an input a set of nodes $V$ with residual degrees $\mathsf{deg}_r(v) > 0, \forall v \in V$ and generates the missing edges degree-wise in a cyclic order. Let $v_{j,1}, \ldots, v_{j,n_j}$ be the nodes of degree $\mathsf{deg}_{\alpha,\beta}(v_{j,l}) = j$ in the set $V$, then the following invariant will be maintained. Since $X \subseteq [x\Delta, y\Delta]$ and $x\Delta$ and $y\Delta$ are chosen such that number of edges $\mathrm{vol}([x\Delta, y\Delta]) = \sum_{j=x\Delta}^{\Delta} \left\lfloor \frac{e^\alpha}{j^\beta} \right\rfloor \cdot j$ (minimally) exceeds the number of nodes in $V_{\alpha,\beta} \setminus X$ we have residual degrees on some $v \in X$ and call `Fill_Wheel`$(X)$. Furthermore, we call `Fill_Wheel`$(W)$ since we have that all $w \in W$ are connected only via a single edge to the set $X$ and $w \in W$ were chosen to have residual degrees in the interval $[3, \Delta]$.

**Invariant 1.** *In every stage of the construction, for every $j \in \{1, \ldots, \Delta\}, \mathsf{deg}_r(v_{j,1}) \leqslant \ldots \leqslant \mathsf{deg}_r(v_{j,n_j})$ and $\mathsf{deg}_r(v_{j,n_j}) - \mathsf{deg}_r(v_{j,1}) \leqslant 1$.*

Figure 2 shows how intervals of uniform residual degrees are filled and how the problems of uneven interval-lengths and uneven residual degrees are resolved at the borders of the intervals.

Depending on the parameter $\beta$, we will show how to choose $x$ and $y$ in such a way that $X$ becomes sufficiently small. Hence any dominating set $D'$ in $\mathcal{G}_{\alpha,\beta}$ can be efficiently transformed

**Algorithm 1:** `ConstructPLG`

**Input**: $G_{U,\mathcal{S}} = (V_{U,\mathcal{S}}, E_{U,\mathcal{S}})$ with $|V_{U,\mathcal{S}}| = N$.
**Output**: Power law graph $\mathcal{G}_{\alpha,\beta} = (V_{\alpha,\beta}, E_{\alpha,\beta})$ with $V_{\alpha,\beta} = V_{U,\mathcal{S}} \cup X \cup W \cup V_1 \cup V_2$, $|V_{\alpha,\beta}| = n$ and $E_{U,\mathcal{S}} \subseteq E_{\alpha,\beta}$.

**choose** $\alpha, x, y$ such that $|[x\Delta, y\Delta]| \geqslant n$ and $|[N^a, N^b]| \geqslant N$;
**set** $X := [x\Delta, y\Delta]$, $W := [3, \Delta] \setminus (V_{U,\mathcal{S}} \cup X)$ and $\Gamma := \emptyset$;
**set** $V_{\alpha,\beta} := V_{U,\mathcal{S}} \cup X \cup W \cup V_1 \cup V_2$;
**for** $i = 1, \ldots, N (= |V_{U,\mathcal{S}}|)$ **do**
    **map** $s_i \in V_{U,\mathcal{S}}$ with $t_i \in V_2 \setminus \Gamma$ and **set** $E_{\alpha,\beta} := E_{\alpha,\beta} \cup \{s_i, t_i\}, \Gamma := \Gamma \cup \{t_i\}$;
    **choose** $v \in X$ with maximum $\mathsf{deg}_r(v) > 0$ and **set** $E_{\alpha,\beta} := E_{\alpha,\beta} \cup \{t_i, v\}$;
    **update** $\mathsf{deg}_r(t_i)$ and $\mathsf{deg}_r(v)$;
**foreach** $u \in V_1 \cup V_2, \mathsf{deg}_r(u) > 0$ **do**
    **choose** $v \in X$ with maximum $\mathsf{deg}_r(v) > 0$ and **set** $E_{\alpha,\beta} := E_{\alpha,\beta} \cup \{u, v\}$;
    **update** $\mathsf{deg}_r(t)$ and $\mathsf{deg}_r(v)$;
**foreach** $w \in W$ **do**
    **choose** $v \in X$ with maximum $\mathsf{deg}_r(v) > 0$ and **set** $E_{\alpha,\beta} := E_{\alpha,\beta} \cup \{w, v\}$;
    **update** $\mathsf{deg}_r(w)$ and $\mathsf{deg}_r(v)$;
`Fill_Wheel`$(W)$;    `/* realizes residual degrees on` $W$ `and` $X$ `*/`
`Fill_Wheel`$(X)$;
**return** $\mathcal{G}_{\alpha,\beta} = (V_{\alpha,\beta}, E_{\alpha,\beta})$;

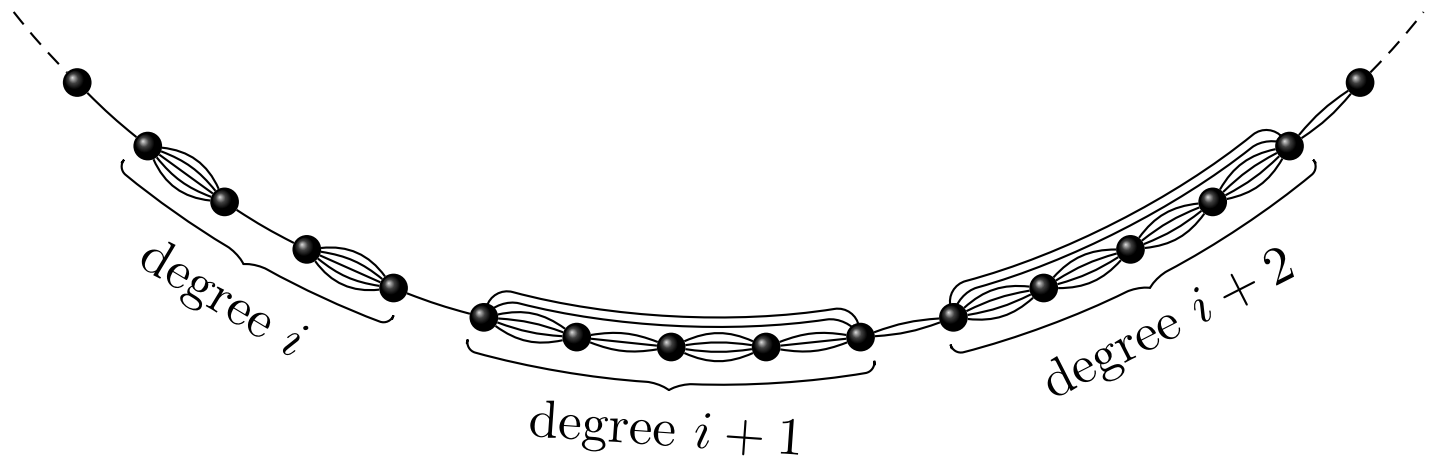

**Figure 2:** Procedure `Fill_Wheel` realizes the residual degrees on the wheel nodes in $W$ and $X$.

into a dominating set $D$ of size $|D| \leqslant |D'|$ such that $D = D_{U,\mathcal{S}} \cup X$, where $D_{U,\mathcal{S}} \subseteq V_{U,\mathcal{S}}$ is a dominating set of $G_{U,\mathcal{S}}$.

# 5 Feige's Lower Bound for Set Cover

Our starting point is U.Feige's logarithmic lower bound for the approximability of the Set Cover problem [Fei98]. For each Set Cover instance $(U, \mathcal{S})$ we embed the associated Minimum Dominating Set instance $G_{U,\mathcal{S}}$ into an $(\alpha, \beta)$-PLG $\mathcal{G}_{\alpha,\beta}$. In order to implement the power law node-degree distribution, we need to know the degree distribution of the graph $G_{U,\mathcal{S}}$. Therefore we briefly review Feige's construction. Feige constructs a $k$-prover proof system for the problem 5OCC-MAX-E3SAT. Consider a 3CNF formula $\varphi$ with $n$ variables such that each variable occurs at most 5 times in $\varphi$. One can assume that either the formula is satisfiable, or no assignment satisfies more than an $\varepsilon$ fraction of the clauses simultaneously. The $k$-prover proof system works as follows: It chooses $k$ codewords of length $l = \Theta(\log \log n)$, weight $\frac{l}{2}$ and pairwise *Hamming distance* $\geqslant \frac{l}{3}$. The verifier picks $l$ clauses $C_1, \ldots, C_l$ from $\varphi$ independently uniformly at random. Independently, from each such clause $C_i$ it picks one variable $x_i$ of $C_i$ uniformly at random. For each $1 \leqslant i \leqslant k$, the verifier sends to the prover $i$ those $\frac{l}{2}$ clauses $C_j$ for which the associated bit of prover $i$'s codeword is 1 and those $\frac{l}{2}$ variables $x_j$ for which the associated bit of prover $i$'s codeword is 0. The provers return their answers, and based on this the verifier determines its output. The construction of the associated Set Cover instances makes use of some combinatorial building blocks called *partition systems*.

According to Feige [Fei98], a partition system $B(m, L, k, d)$ consists of a ground set $B$ of cardinality $|B| = m$ and $L$ partitions $p_1, \ldots, p_L$ of $B$ into $k$ disjoint subsets $p_{j,h} \subset B$. The defining property of these partition systems is that each cover of $B$ by subsets $p_{j,h}$ which uses sets from pairwise different partitions must consist of at least $d$ subsets. Feige gives a randomized construction of such partition systems with $L \approx (\log m)^c$, $k$ being any number smaller than $\ln(\frac{m}{3}) \cdot \ln\ln(m)$ and $d = (1 - f(k)) \cdot k \cdot \ln(m)$ with some function $f(k)$ with $f(k) \longrightarrow 0$ as $k \longrightarrow \infty$. That construction yields partitions for which with high probability all the sets have the same size. We show that the same result is obtained by making use of random permutations. But now, for each partition $p_j$, the sets $p_{j,h}$ always have the same size $\frac{m}{k}$ (provided $k|m$). Namely, choose a random permutation $\pi_j \in_R S_m$ and let $p_{j,h} = \{\pi_j((h-1)\frac{m}{k}+1), \ldots, \pi_j(k \cdot \frac{m}{k})\}$. Suppose now we cover $B$ with $d$ subsets $p_{j_1,h_1}, \ldots, p_{j_d,h_d}$ from pairwise different partitions. Then for a given point $v \in B$, the probability that $v$ is covered by at least one of them is

$$\begin{aligned}
&P(\text{point } v \in B \text{ is covered by at least one of these } d \text{ sets})\\
&= 1 - \prod_{i=1}^{d} P\left(v \text{ is not in position } 1, \ldots, \frac{m}{k} \text{ in permutation } \pi_j\right)\\
&= 1 - \left(\frac{\binom{m-1}{m/k} \cdot (\frac{m}{k})! \cdot (m - \frac{m}{k})!}{m!}\right)^d\\
&= 1 - \left(\frac{(m-1)! \cdot (m - \frac{m}{k})!}{(m - 1 - \frac{m}{k})! \cdot m!}\right)^d\\
&= 1 - \left(\frac{m \cdot \left(1 - \frac{1}{k}\right)}{m}\right)^d = 1 - \left(1 - \frac{1}{k}\right)^d
\end{aligned}$$

This is precisely the property of the randomized construction which has been used by Feige in the analysis of the construction. So from now on we assume that all sets of a partition $p_j$ have the same size $\frac{m}{k}$.

**Feige's Set Cover Instances.** For a given 5OCC-MAX-E3SAT formula $\varphi$ with $n$ variables and the property that either $\varphi$ is satisfiable or no assignment satisfies more than an $\varepsilon$ fraction of the clauses, Feige constructs a Set Cover instance $U, \mathcal{S}$ as follows:

- $\mathcal{R}$ is the set of random strings used by the verifier in the $k$-Prover Proof System. The number of random strings is $|\mathcal{R}| = R = (5n)^l$.
- $|U| = mR$ with $m = (5n)^{\frac{2l}{\varepsilon}}$, hence $|U| = (5n)^{l\left(1+\frac{2}{\varepsilon}\right)}$
- For each $r \in \mathcal{R}$, $B_r(m, L, k, d)$ is a partition system with $L = 2^l$.
- $Q = n^{\frac{l}{2}} \cdot \left(\frac{5n}{3}\right)^{\frac{l}{2}}$ is the number of different queries the verifier may ask to a prover.
- $\mathcal{S}$ contains for every triple $(q, a, i)$ a set $S_{q,a,i}$, where $q$ is a query, $i$ is (the index of) a prover and $a$ is the prover's answer. The set $S_{q,a,i}$ is defined as $S_{q,a,i} = \bigcup_{r:\,(q,i)\in r} B(r, a_r, i)$.

Hence the number of sets in $\mathcal{S}$ is $Q \cdot k$, and each set is of cardinality $\sqrt{R} \cdot \frac{m}{k}$. In how many sets does a point (an element of $U$) occur? For each prover $i$, for each query $q$, each point in $B_r$ with $|B_r| = m$ occurs in $2^l$ sets $S_{q,a,i}$. Hence the total degree of points (= #occurrences of this point in sets) is $2^l \cdot Q$.

**From Set Cover to Dominating Set.** Let $(U, \mathcal{S})$ denote a Set Cover instance with $U = \{u_1, \ldots, u_{|U|}\}$ and $\mathcal{S} = \left\{S_1, \ldots, S_{|\mathcal{S}|}\right\}$. Let $G_{U,\mathcal{S}}$ be the undirected graph with set of vertices $V_{U,\mathcal{S}} = U \cup \mathcal{S}$ and set of edges $E_{U,\mathcal{S}} = \{\{S_i, u_j\} | u_j \in S_i\} \cup \{\{S_j, S_l\} | \, S_j \cap S_l \neq \emptyset\}$. We observe that each set cover $\mathcal{C} \subseteq \mathcal{S}$ is a dominating set in $G_{U,\mathcal{S}}$. On the other hand, let $D \subseteq V_{U,\mathcal{S}}$ be a dominating set in $G_{U,\mathcal{S}}$ with $D = D_U \cup D_{\mathcal{S}}, D_U = D \cap U$ and $D_{\mathcal{S}} = D \cap \mathcal{S}$. If we replace each $u_i \in D_U$ by an arbitrary set $S_j$ with $u_i \in S_j$, the resulting set $D'$ is a dominating set with $D_{\mathcal{S}} \subseteq D' \subseteq \mathcal{S}$ and $|D'| \leqslant |D|$. Hence in this sense we can say dominating sets in $G_{U,\mathcal{S}}$ correspond to set covers $\mathcal{C}$ for $U, \mathcal{S}$.

In Feige's construction, the parameter $l$ satisfies $l = \Theta(\log \log n)$. If $N_0 = |U| + |\mathcal{S}|$ is the number of nodes of $G_{U,\mathcal{S}}$, then (up to logarithmic factors), $N_0 \approx n^l + n^{l\left(1+\frac{2}{\varepsilon}\right)}$, the degree of element nodes $u \in U$ is $\approx n^l$, each set contains $n^{l\left(\frac{1}{2}+\frac{2}{\varepsilon}\right)}$ elements and there are $\approx n^l$ sets. The degree of set nodes in $G_{U,\mathcal{S}}$ is bounded by the sum of the cardinality of that set and the number of sets in the instance $U, \mathcal{S}$, which is $\approx n^{l\left(\frac{1}{2}+\frac{2}{\varepsilon}\right)}$. Hence we obtain the following result.

**Lemma 1.** *Let $F_{SC}$ denote Feige's reduction from 5OCC-MAX-E3SAT to the* Set Cover *problem, and for a given* Set Cover *instance $U, \mathcal{S} = f(\varphi)$ let $G_{U,\mathcal{S}}$ be the associated* Min-DS *instance as described above. If $N_0$ is the number of nodes of $G_{U,\mathcal{S}}$, then for every node $v$ in $G_{U,\mathcal{S}}$, the node degree of $v$ in $G_{U,\mathcal{S}}$ satisfies $N_0^a \leqslant \mathsf{deg}_{U,\mathcal{S}}(v) \leqslant N_0^b$, where $0 < a < b < 1$ and*

$$b \;=\; (1 + o(1)) \cdot \frac{\frac{1}{2} + \frac{2}{\varepsilon}}{1 + \frac{2}{\varepsilon}} \;=\; (1 + o(1)) \cdot \frac{\varepsilon + 4}{2\varepsilon + 4}$$

# 6 New Lower Bounds

We will now describe our new logarithmic lower bounds for approximability of the Minimum Dominating Set problem in $(\alpha, \beta)$-PLG. We distinguish several cases depending on the range of the parameter $\beta$. For the case $1 < \beta < 2$ our construction involves rescaling of the instances $G_{U,\mathcal{S}}$, which has the effect of shifting the degree interval $\left[N^a, N^b\right]$ towards the left end of the full interval $[1, \Delta]$. It turns out that for the case $\beta = 1$ we can omit the scaling and directly implement the power law distribution.

## 6.1 The Case $1 < \beta < 2$

We consider the case $1 < \beta < 2$. Let $(U, \mathcal{S})$ be an instance of the Set Cover problem which is an image $(U, \mathcal{S}) = F_{SC}(\varphi, \varepsilon)$ of some 5OCC-MAX-E3SAT instance $\varphi$ under Feige's reduction $F_{SC}$ with parameter $\varepsilon > 0$. Suppose the number of nodes of $G_{U,\mathcal{S}}$ is $N_0$. Let $\mathsf{OPT}(G_{U,\mathcal{S}})$ denote a minimum cardinality dominating set of $G_{U,\mathcal{S}}$. Then

$$|\mathsf{OPT}(G_{U,\mathcal{S}})| \leqslant k \cdot N_0^{\frac{\varepsilon}{2+\varepsilon}}$$

or

$$|\mathsf{OPT}(G_{U,\mathcal{S}})| \geqslant (1 - \varepsilon) \cdot k \cdot N_0^{\frac{\varepsilon}{2+\varepsilon}} \cdot \frac{\varepsilon}{2 + \varepsilon} \cdot \left(\frac{1}{2}\right)^{\frac{\varepsilon}{2+\varepsilon}} \cdot (\ln(N_0) - O(1)),$$

where $k$ is the number of provers in Feige's $k$-prover proof system. Furthermore, the node degrees in $G_{U,\mathcal{S}}$ are contained in the interval $\left[N_0^a, N_0^b\right]$ with $0 < a < b < 1$ being constant.

**Scaling.** In the case $1 < \beta < 2$, it turns out that we have to rescale the node degrees of nodes in $G_{U,\mathcal{S}}$ appropriately. Namely, we replace $G_{U,\mathcal{S}}$ by the graph $G^d_{U,\mathcal{S}}$ which consists of $N_0^{d-1}$ disjoint copies of the graph $G_{U,\mathcal{S}}$. Here, $d$ is a parameter of our construction. The graph $G^d_{U,\mathcal{S}}$ has the following properties:

- The number of nodes is $N := N_0^d$.
- The node degrees are contained in the interval $\left[N^{a/d}, N^{b/d}\right]$
- Let $\mathsf{OPT}(G^d_{U,\mathcal{S}})$ denote an optimum dominating set of $G_{U,\mathcal{S}}$. Then

$$|\mathsf{OPT}(G^d_{U,\mathcal{S}})| \leqslant N^{\frac{d-1}{d}} \cdot k \cdot N^{\frac{1}{d}\cdot\frac{\varepsilon}{2+\varepsilon}} = k \cdot N^{\frac{1}{d}\cdot\left(d-1+\frac{\varepsilon}{2+\varepsilon}\right)}$$

or

$$\begin{aligned}|\mathsf{OPT}(G^d_{U,\mathcal{S}})| &\geqslant (1-\varepsilon)\cdot k \cdot N^{\frac{1}{d}\cdot\frac{\varepsilon}{2+\varepsilon}} \cdot \frac{\varepsilon}{2+\varepsilon} \cdot \left(\frac{1}{2}\right)^{\frac{\varepsilon}{2+\varepsilon}} \cdot \left(\ln\left(N^{\frac{1}{d}}\right) - O(1)\right) \cdot N^{\frac{d-1}{d}} \\ &= k \cdot \frac{\varepsilon(1-\varepsilon)}{2+\varepsilon} \cdot \left(\frac{1}{2}\right)^{\frac{\varepsilon}{2+\varepsilon}} \cdot N^{\frac{1}{d}\cdot\left(d-1+\frac{\varepsilon}{2+\varepsilon}\right)} \cdot \left(\ln\left(N^{\frac{1}{d}}\right) - O(1)\right)\end{aligned}$$

**Construction of $\mathcal{G}_{\alpha,\beta}$.** We choose $\alpha$ and the parameters $x, y$ such as to satisfy the following constraints:

1. $\left|\left[N^{a/d}, N^{b/d}\right]\right| \geqslant N$
2. $|[x\Delta, y\Delta]| = o\left(N^{\frac{d-1}{d}}\right)$, where $N^{\frac{d-1}{d}}$ is a lower bound for the size of an optimum dominating set in $G_{U,\mathcal{S}}$.
3. $\sum_{j=x\Delta}^{y\Delta} \lfloor \frac{e^\alpha}{j^\beta} \rfloor \cdot j = \mathrm{vol}(|x\Delta, y\Delta|) \geqslant \zeta(\beta)\cdot e^\alpha$
(the total node degree of the set $[x\Delta, y\Delta]$ is large enough such that $[x\Delta, y\Delta]$ can dominate the wheel $W$ as well as all the degree-2 nodes which are matched to nodes in $G_{U,\mathcal{S}}$)

Constraint (1) is implied by the following stronger constraint (1').

1'. $\frac{e^\alpha}{N^{\frac{b\beta}{d}}} \geqslant N$

We work with (1') instead of (1) and obtain the following bound for $\alpha$:

$$e^\alpha \geqslant N^{1+\frac{b\beta}{d}}$$

Thus we choose $e^\alpha = N^{1+\frac{b\beta}{d}}$. We have

$$\begin{aligned}|[x\Delta, y\Delta]| &= \sum_{x\Delta}^{y\Delta} \left\lfloor \frac{e^\alpha}{j^\beta} \right\rfloor \in \left[\frac{e^\alpha}{\Delta^\beta}\cdot(y-x)\cdot\Delta\cdot\frac{1}{y^\beta} - (y-x)\Delta, \frac{e^\alpha}{\Delta^\beta}\cdot(y-x)\cdot\Delta\cdot\frac{1}{x^\beta}\right] \\ &= \left[\Delta(y-x)\left(\frac{1}{y^\beta}-1\right), \Delta\cdot\frac{y-x}{x^\beta}\right]\end{aligned}$$

and for the volume of the interval $\mathrm{vol}(|x\Delta, y\Delta|) = \sum_{j=x\Delta}^{y\Delta} \left\lfloor \frac{e^\alpha}{j^\beta} \right\rfloor \cdot j$

$$\begin{aligned}
\mathrm{vol}(|x\Delta,y\Delta|) &\geqslant e^\alpha \cdot \sum_{j=x\Delta}^{y\Delta} j^{1-\beta} - r_\beta = (1-o(1))e^\alpha \cdot \int_{x\Delta}^{y\Delta} j^{1-\beta}\,\mathrm{d}j - r_\beta \\
&= (1-o(1))e^\alpha \cdot \left[\frac{j^{2-\beta}}{2-\beta}\right]_{x\Delta}^{y\Delta} - r_\beta = (1-o(1))e^\alpha \cdot e^{\alpha\cdot\frac{2-\beta}{\beta}} \cdot \frac{y^{2-\beta}-x^{2-\beta}}{2-\beta} - r_\beta \\
&= (1-o(1))\Delta^2 \cdot \frac{y^{2-\beta}-x^{2-\beta}}{2-\beta} - r_\beta
\end{aligned}$$

where $r_\beta = \frac{\Delta^2(y^2)-x^2)}{2} + \frac{\Delta(y+x)}{2}$ is an upper bound for the rounding error. Hence we obtain $\mathrm{vol}([x\Delta, y\Delta]) = \omega(|\mathcal{G}_{\alpha,\beta}|)$ provided we choose $x$ and $y$ in such a way that $\frac{y^{2-\beta}-x^{2-\beta}}{2-\beta} - r_\beta > 0$. Let us choose $y = 1$. Then we have

$$\frac{y^{2-\beta}-x^{2-\beta}}{2-\beta} - r_\beta = \frac{1-x^{2-\beta}}{2-\beta} - \frac{1-x^2}{2} - o(1) = \frac{\beta - 2x^{2-\beta} + (2-\beta)x^2}{2\cdot(2-\beta)} - o(1)$$

Hence we want to choose $x \in (0,1)$ such that $\beta - 2x^{2-\beta} + (2-\beta)x^2 > 0$. This inequality holds for $x < \left(\frac{\beta}{2}\right)^{\frac{1}{2-\beta}}$, since $\frac{\beta}{2} < 1$.

For our choice of $\alpha$, $N^{\frac{d-1}{d}} = e^{\alpha\cdot\frac{d-1}{d+b\beta}}$, and hence constraint (2) holds if the following constraint is satisfied:

$$\Delta \cdot \frac{y-x}{x^\beta} = \frac{y-x}{x^\beta} \cdot e^{\frac{\alpha}{\beta}} = o\left(e^{\alpha\cdot\frac{d-1}{d+b\beta}}\right)$$

Hence for our choice of $y = 1, x < \left(\frac{\beta}{2}\right)^{\frac{1}{2-\beta}}$ this last constraint is satisfied if $\frac{\alpha}{\beta} < \alpha \cdot \frac{d-1}{d+b\beta}$, i.e. $d > \frac{(b+1)\beta}{\beta-1}$.

**Resulting Lower Bound.** Since $e^\alpha = N^{1+\frac{b\beta}{d}}$, we have $|\mathcal{G}_{\alpha,\beta}| = \zeta(\beta)\cdot N^{1+\frac{b\beta}{d}}$ and thus obtain the following bounds for the size of an optimum dominating set for $\mathcal{G}_{\alpha,\beta}$:

$$|\mathsf{OPT}(\mathcal{G}_{\alpha,\beta})| \leqslant \left(\left(\frac{|\mathcal{G}_{\alpha,\beta}|}{\zeta(\beta)}\right)^{\frac{d}{d+b\beta}}\right)^{\frac{d-1}{d}} \cdot k \cdot \left(\left(\frac{|\mathcal{G}_{\alpha,\beta}|}{\zeta(\beta)}\right)^{\frac{d}{d+b\beta}}\right)^{\frac{1}{d}\cdot\frac{\varepsilon}{2+\varepsilon}} = k \cdot \left(\frac{|\mathcal{G}_{\alpha,\beta}|}{\zeta(\beta)}\right)^{\frac{d-1+\frac{\varepsilon}{2+\varepsilon}}{d+b\beta}}$$

or

$$|\mathsf{OPT}(\mathcal{G}_{\alpha,\beta})| \geqslant k \cdot \left(\frac{|\mathcal{G}_{\alpha,\beta}|}{\zeta(\beta)}\right)^{\frac{d-1+\frac{\varepsilon}{2+\varepsilon}}{d+b\beta}} \cdot \frac{(1-\varepsilon)\cdot\varepsilon}{2+\varepsilon}\left(\frac{1}{2}\right)^{\frac{\varepsilon}{2+\varepsilon}} \cdot \left(\ln\left(\left(\frac{|\mathcal{G}_{\alpha,\beta}|}{\zeta(\beta)}\right)^{\frac{d}{d+b\beta}\cdot\frac{1}{d}}\right) - O(1)\right)$$

Altogether, we obtain the following theorem.

**Theorem 1.** *For $1 < \beta < 2$, the* Min-DS *problem on $(\alpha,\beta)$-Power Law Graphs is hard to approximate within*

$$\frac{(1-\varepsilon)\cdot\varepsilon}{2+\varepsilon} \cdot \left(\frac{1}{2}\right)^{\frac{\varepsilon}{2+\varepsilon}} \cdot \frac{\ln(|\mathcal{G}_{\alpha,\beta}|) - \ln(\zeta(\beta))}{d+b\beta}$$

## 6.2 The Case $0 < \beta < 1$

Let us now consider the case $0 < \beta < 1$. We will again have to make use of Scaling. Furthermore, in this case we have to choose parameters $x, y$ of the interval $X = [x\Delta, y\Delta]$ carefully in order to obtain a logarithmic lower bound. First we give an estimate for the size of the interval $[x\Delta, y\Delta]$ and the sum of node degrees of nodes in this interval. We have

$$|[x\Delta, y\Delta]| \in \left[\sum_{j=x\Delta}^{y\Delta} \frac{e^\alpha}{j^\beta} - (y-x+1)\Delta,\ \sum_{j=x\Delta}^{y\Delta} \frac{e^\alpha}{j^\beta}\right]$$

where

$$\begin{aligned}
\sum_{j=x\Delta}^{y\Delta} \frac{e^\alpha}{j^\beta} &\in \left[e^\alpha \int_{x\Delta}^{y\Delta} \frac{1}{j^\beta}\,\mathrm{d}j - e^\alpha\left(\frac{1}{(x\Delta)^\beta} - \frac{1}{(y\Delta)^\beta}\right),\ e^\alpha \int_{x\Delta}^{y\Delta} \frac{1}{j^\beta}\,\mathrm{d}j\right] \\
&= \left[e^\alpha \left[\frac{j^{1-\beta}}{1-\beta}\right]_{x\Delta}^{y\Delta} - \frac{e^\alpha}{\Delta^\beta}\left(\frac{1}{x^\beta} - \frac{1}{y^\beta}\right),\ e^\alpha \left[\frac{j^{1-\beta}}{1-\beta}\right]_{x\Delta}^{y\Delta}\right] \\
&= \left[\frac{e^\alpha \Delta^{1-\beta}}{1-\beta}\left(y^{1-\beta} - x^{1-\beta}\right) - \left(\frac{1}{x^\beta} - \frac{1}{y^\beta}\right),\ \frac{e^\alpha \Delta^{1-\beta}}{1-\beta}\left(y^{1-\beta} - x^{1-\beta}\right)\right] \\
&= \left[\frac{\Delta}{1-\beta}\left(y^{1-\beta} - x^{1-\beta}\right) - \left(\frac{1}{x^\beta} - \frac{1}{y^\beta}\right),\ \frac{\Delta}{1-\beta}\left(y^{1-\beta} - x^{1-\beta}\right)\right]
\end{aligned}$$

The sum of node degrees of nodes in $[x\Delta, y\Delta]$ is

$$\mathrm{vol}([x\Delta, y\Delta]) = \sum_{x\Delta}^{y\Delta} \left\lfloor \frac{e^\alpha}{j^\beta} \right\rfloor \cdot j \in \left[\sum_{x\Delta}^{y\Delta} \frac{e^\alpha}{j^{\beta-1}} - \underbrace{\left(\frac{y\Delta(y\Delta-1)}{2} - \frac{x\Delta(x\Delta-1)}{2}\right)}_{\text{rounding error}},\ \sum_{x\Delta}^{y\Delta} \frac{e^\alpha}{j^{\beta-1}}\right]$$

where

$$\begin{aligned}
\sum_{x\Delta}^{y\Delta} \frac{e^\alpha}{j^{\beta-1}} &\in \left[e^\alpha \int_{x\Delta}^{y\Delta} j^{1-\beta}\,\mathrm{d}j - e^\alpha\left((y\Delta)^{1-\beta} - (x\Delta)^{1-\beta}\right),\ e^\alpha \int_{x\Delta}^{y\Delta} j^{1-\beta}\,\mathrm{d}j\right] \\
&= \left[e^\alpha \left[\frac{j^{2-\beta}}{2-\beta}\right]_{x\Delta}^{y\Delta} - \Delta\left(y^{1-\beta} - x^{1-\beta}\right),\ e^\alpha \left[\frac{j^{2-\beta}}{2-\beta}\right]_{x\Delta}^{y\Delta}\right] \\
&= \left[\frac{\Delta^2}{2-\beta}\left(y^{2-\beta} - x^{2-\beta}\right) - \Delta\left(y^{1-\beta} - x^{1-\beta}\right),\ \frac{\Delta^2}{2-\beta}\left(y^{2-\beta} - x^{2-\beta}\right)\right]
\end{aligned}$$

We choose $y = 1$ and obtain

$$|[x\Delta, \Delta]| \in \left[\frac{\Delta}{1-\beta}\left(1 - x^{1-\beta}\right) - \left(\frac{1}{x^\beta} - 1\right) - (2-x)\Delta,\ \frac{\Delta}{1-\beta}\left(1 - x^{1-\beta}\right)\right]$$

The volume of that interval is then estimated as

$$\begin{aligned}
\mathrm{vol}([x\Delta, \Delta]) &\geqslant \frac{\Delta^2}{2-\beta}\left(1 - x^{2-\beta}\right) - \Delta\left(1 - x^{1-\beta}\right) - \left(\frac{\Delta(\Delta+1)}{2} - \frac{x^2\Delta^2 - x\Delta}{2}\right) \\
&= \frac{\Delta^2}{2-\beta}\left(1 - x^{2-\beta}\right) - \frac{\Delta^2}{2} + \frac{x^2}{2}\Delta^2 - \Delta\left(1 - x^{1-\beta} - \frac{1}{2} + \frac{x}{2}\right) \\
&= \Delta^2\left(\frac{1 - x^{2-\beta}}{2-\beta} - \frac{1}{2} + \frac{x^2}{2}\right) - \Delta\left(1 - x^{1-\beta} - \frac{1}{2} + \frac{x}{2}\right)
\end{aligned}$$

We use *Scaling* with the scaling parameter $d$, hence we want to choose $\alpha$ such that $e^\alpha \geqslant N^{\frac{d+b\beta}{d}}$. Since $N^{\frac{d-1}{d}}$ is a lower bound for the optimum in $G^d_{U,\mathcal{S}}$, we have $N^{\frac{d-1}{d}} = e^{\frac{d-1}{d+b\beta}\cdot\alpha} = e^{(1-\delta)\alpha}$, where we can choose $1-\delta$ arbitrary close to 1. The size of the interval $[x\Delta, \Delta]$ is of order $\Delta(1-x^{1-\beta})$, hence we want to choose $x$ such that $\Delta(1-x^{1-\beta}) = e^{\alpha/\beta}\cdot e^p$ with $\frac{\alpha}{\beta}\cdot p < (1-\delta)\alpha$, i.e. $p < (1-\delta)\beta$. So suppose we choose $x$ such that $p = (1-\delta')\beta$, where $1-\delta'$ can be chosen arbitrary close to 1. Furthermore, the interval $[x\Delta, \Delta]$ needs to provide sufficient volume to dominate the rest of the graph, i.e. (using our volume estimate)

$$\Delta^2\left(\frac{1}{2-\beta} - \frac{1}{2} - x^{2-\beta}\left(\frac{1}{2-\beta} - \frac{x^\beta}{2}\right)\right) > \Delta$$

This yields the requirement $\frac{1}{2-\beta} - \frac{1}{2} - x^{2-\beta}\left(\frac{1}{2-\beta} - \frac{x^\beta}{2}\right) > \frac{1}{\Delta}$, which is implied by

$$1 - \frac{1}{\Delta\left(\frac{1}{2-\beta} - \frac{1}{2}\right)} > x^2$$

Combining this with the upper bound requirement for the size of the interval, we obtain

$$\left(1 - \frac{1-\beta}{e^{\alpha\left(\frac{1}{\beta}-(1-\delta')\right)}}\right)^{\frac{1}{1-\beta}} \leqslant x < \left(1 - \frac{1}{\left(\frac{1}{2-\beta} - \frac{1}{2}\right)\cdot e^{\alpha/\beta}}\right)^{1/2} \tag{1}$$

We observe that $\frac{1}{1-\beta} > 1 > \frac{1}{2}$ for $\beta \in (0,1)$, and furthermore $\frac{\alpha}{\beta} - (1-\delta')\alpha < \frac{\alpha}{\beta}$, hence we can choose $x$ such that Equation 1 holds. Thus for this choice of $x$ we have $|[x\Delta, \Delta]| = o\left(N^{\frac{d-1}{d}}\right)$ and $\mathrm{vol}([x\Delta, \Delta]) \geqslant |\mathcal{G}_{\alpha,\beta}|$. As in the case $1 < \beta < 2$, we have $\mathsf{OPT}(\mathcal{G}_{\alpha,\beta}) = (1+o(1))\mathsf{OPT}(G^d_{U,\mathcal{S}})$, and furthermore $N = (|\mathcal{G}_{\alpha,\beta}| \cdot (1-\beta))^{\frac{d\beta}{d+b\beta}}$. Altogether we obtain the following result.

**Theorem 2.** *For $0 < \beta < 1$, the* Min-DS *problem on $(\alpha,\beta)$-Power Law Graphs is hard to approximate within*

$$\frac{(1-\varepsilon)\varepsilon}{2+\varepsilon}\cdot\left(\frac{1}{2}\right)^{\frac{\varepsilon}{2+\varepsilon}}\cdot\left(\frac{\beta}{d+b\beta}\cdot\left(\ln(|\mathcal{G}_{\alpha,\beta}|) - \ln\left(\frac{1}{1-\beta}\right)\right) - O(1)\right)$$

### 6.3 The Case $\beta = 1$

In the case $\beta = 1$ we can omit the scaling and directly embed the graph $G_{U,\mathcal{S}}$ into a PLG $\mathcal{G}_{\alpha,\beta}$. It suffices to describe the choice of parameters $x, \alpha$ for a given $G_{U,\mathcal{S}}$ and to verify that the requirements (1)-(3) are satisfied. For a given $x \in [0,1]$, the size of the interval $[x\Delta, \Delta] = \{v \in V(\mathcal{G}_{\alpha,\beta}) \mid x\Delta \leqslant \mathsf{deg}_{\alpha,\beta}(v) \leqslant \Delta\}$ satisfies

$$\begin{aligned}
|[x\Delta, \Delta]| = \sum_{x\Delta}^{\Delta}\left\lfloor\frac{e^\alpha}{j}\right\rfloor &\in \left[\sum_{xe^\alpha}^{e^\alpha}\frac{e^\alpha}{j} - (1-x)e^\alpha,\ \sum_{xe^\alpha}^{e^\alpha}\frac{e^\alpha}{j}\right] \\
&\subseteq \left[e^\alpha\left(\ln(e^\alpha) - \ln(xe^\alpha)\right) - e^\alpha(\frac{1}{x}-1)\cdot\frac{1}{e^\alpha},\ e^\alpha\cdot\ln\left(\frac{1}{x}\right)\right] \\
&= \left[e^\alpha\ln\left(\frac{1}{x}\right) - \left(\frac{1}{x}-1\right),\ e^\alpha\ln\left(\frac{1}{x}\right)\right]
\end{aligned}$$

The volume of that interval is

$$\begin{aligned}\mathrm{vol}([x\Delta,\Delta]) &= \sum_{x\Delta}^{\Delta}\left\lfloor\frac{e^\alpha}{j}\right\rfloor\cdot j \;\in\; \left[\sum_{x\Delta}^{\Delta}e^\alpha \;-\; j,\; \sum_{x\Delta}^{\Delta}e^\alpha\right] \\ &\subseteq \left[e^\alpha(1-x)\Delta \;-\; \left(\frac{\Delta(\Delta+1)}{2}-\frac{x\Delta(x\Delta+1)}{2}\right),\; e^\alpha(1-x)\Delta\right] \\ &= \left[\Delta^2\left(\frac{1}{2}-x+\frac{x^2}{2}\right) \;-\; \frac{1-x}{2}\Delta,\; (1-x)\Delta^2\right]\end{aligned}$$

Hence for every $x<1$ being bounded away from 1, the volume of the interval $[x\Delta,\Delta]$ is $\omega(|\mathcal{G}_{\alpha,1}|)$. Recall that in order to achieve $N_0 \leqslant \left|\left[N_0^a, N_0^b\right]\right|$, it suffices to choose $\alpha$ sufficiently large such that $N_0 \leqslant \frac{e^\alpha}{N_0^{b\beta}} = \frac{e^\alpha}{N_0^b}$. Hence suppose we have $N_0^{1+b} = e^\alpha$. This implies $\frac{e^\alpha}{N_0^b} = e^{\alpha\cdot\frac{1}{1+b}}$. Thus it suffices to choose $x$ such that $\ln\left(\frac{1}{x}\right) = o\left(e^{\alpha\cdot\frac{b}{1+b}}\right)$.

The size of the PLG is $|\mathcal{G}_{\alpha,\beta}| = \alpha e^\alpha$, and from $N_0^{1+b} = e^\alpha$ we obtain $N_0 = e^{\frac{\alpha}{1+b}} = \left(\frac{|\mathcal{G}_{\alpha,\beta}|}{\ln(\mathcal{G}_{\alpha,\beta})}\right)^{\frac{1}{1+b}}$. Hence we obtain the following lower bound for the case $\beta=1$.

**Theorem 3.** *For $\beta=1$, the* MIN-DS *problem on $(\alpha,\beta)$-Power Law Graphs is hard to approximate within*

$$\frac{(1-\varepsilon)\varepsilon}{2+\varepsilon}\cdot\left(\frac{1}{2}\right)^{\frac{\varepsilon}{2+\varepsilon}}\cdot\left(\frac{(1-o(1))\ln(|\mathcal{G}_{\alpha,\beta}|)}{1+b}-O(1)\right)$$

### 6.4 The Case $\beta = 2$

In this case, again, we give an estimate for the size of the interval $[x\Delta, y\Delta]$ and for the sum of node degrees inside this interval. We have that

$$|[x\Delta,y\Delta]| \in \left[\Delta\frac{y-x}{y^\beta},\Delta\frac{y-x}{x^\beta}\right] \;=\; \left[\sqrt{e^\alpha}\cdot\frac{y-x}{y^\beta},\sqrt{e^\alpha}\cdot\frac{y-x}{x^\beta}\right].$$

The volume $\mathrm{vol}([x\Delta,y\Delta])$ of the interval is $(1-o(1))\sum_{j=x\Delta}^{y\Delta}\frac{e^\alpha}{j^\beta}\cdot j = (1-o(1))e^\alpha\left(\ln(y\Delta)-\ln(x\Delta)\right)$ $= (1-o(1))e^\alpha\left(\ln\left(\frac{1}{x}\right)-\ln\left(\frac{1}{y}\right)\right)$. We choose $y=1$ and obtain

$$\mathrm{vol}([x\Delta,y\Delta]) = (1-o(1))\sum_{j=x\Delta}^{y\Delta}\frac{e^\alpha}{j^\beta}\cdot j = (1-o(1))e^\alpha\left(\ln\left(\frac{1}{x}\right)-0\right).$$

Hence, want to choose $x$ such that $\ln\left(\frac{1}{x}\right) \geqslant \zeta(\beta)$, i.e. $x \leqslant \frac{1}{e^{\zeta(\beta)}}$. Then the volume of the interval $[x\Delta,\Delta]$ suffices to dominate the rest of the graph, hence constraint (3) is satisfied. The size of the interval $[x\Delta,\Delta]$ satisfies $|[x\Delta,\Delta]| \in \left[\Delta\frac{1-x}{1},\Delta\frac{1-x}{x^\beta}\right]$. The two intervals $[x\Delta,\Delta]$ and $[N^{a/d},N^{b/d}]$ need to be node disjoint. Hence we want to choose $d$ such that $N^{b/d} < x\Delta$. For $x = \frac{1}{e^{\zeta(\beta)}}$, we have $x\Delta = e^{\alpha/\beta-\zeta(\beta)}$. Furthermore, the size $N$ of the graph $G^d_{U,\mathcal{S}}$ satisfies $N = |G^d_{U,\mathcal{S}}| \leqslant e^{\alpha\frac{d}{d+b\beta}}$. This yields the following bound for the scaling parameter $d$:

$$N^{b/d} \;<\; x\Delta \iff e^{\alpha\cdot b\cdot\frac{1}{d+b\beta}} \;<\; e^{\alpha/\beta-\zeta(\beta)} \iff d \;>\; \frac{\alpha\cdot b}{\alpha/\beta-\zeta(\beta)} - b\beta.$$

**Resulting Lower Bound.** Constraint (1') yields the following bound for the size of the power law graph: $e^\alpha \geqslant N^{1+\frac{b\beta}{d}}$. Thus we choose $e^\alpha = N^{1+\frac{b\beta}{d}}$ which implies $|\mathcal{G}_{\alpha,\beta}| = \zeta(\beta) \cdot N^{1+\frac{b\beta}{d}}$. Thus we obtain the following bounds for the size of an optimum dominating set for $\mathcal{G}_{\alpha,\beta}$:

$$|\mathsf{OPT}(\mathcal{G}_{\alpha,\beta})| \leqslant \left(\left(\frac{|\mathcal{G}_{\alpha,\beta}|}{\zeta(\beta)}\right)^{\frac{d}{d+b\beta}}\right)^{\frac{d-1}{d}} \cdot k \cdot \left(\left(\frac{|\mathcal{G}_{\alpha,\beta}|}{\zeta(\beta)}\right)^{\frac{d}{d+b\beta}}\right)^{\frac{1}{d}\cdot\frac{\varepsilon}{2+\varepsilon}} = k \cdot \left(\frac{|\mathcal{G}_{\alpha,\beta}|}{\zeta(\beta)}\right)^{\frac{d-1+\frac{\varepsilon}{2+\varepsilon}}{d+b\beta}}$$

or

$$|\mathsf{OPT}(\mathcal{G}_{\alpha,\beta})| \geqslant k \cdot \left(\frac{|\mathcal{G}_{\alpha,\beta}|}{\zeta(\beta)}\right)^{\frac{d-1+\frac{\varepsilon}{2+\varepsilon}}{d+b\beta}} \cdot \frac{(1-\varepsilon)\cdot\varepsilon}{2+\varepsilon}\left(\frac{1}{2}\right)^{\frac{\varepsilon}{2+\varepsilon}} \cdot \left(\ln\left(\left(\frac{|\mathcal{G}_{\alpha,\beta}|}{\zeta(\beta)}\right)^{\frac{d}{d+b\beta}\cdot\frac{1}{d}}\right) - O(1)\right)$$

Hence, we obtain the following result.

**Theorem 4.** *For $\beta = 2$, the* Min-DS *problem on $(\alpha,\beta)$-Power Law Graphs is hard to approximate within*

$$\frac{(1-\varepsilon)\cdot\varepsilon}{2+\varepsilon} \cdot \left(\frac{1}{2}\right)^{\frac{\varepsilon}{2+\varepsilon}} \cdot \frac{\ln\left(|\mathcal{G}_{\alpha,\beta}|\right) - \ln(\zeta(\beta))}{d+b\beta}$$

# 7 New Upper Bounds for $\beta > 2$

For $\beta > 2$, the Min-DS problem on $(\alpha,\beta)$-PLG is in APX. This was already observed by Shen et al. in [She+12]. They showed that in that case, there exists an efficient approximation algorithm with approximation ratio $(\zeta(\beta) - 1/2)/(\zeta(\beta) - \sum_{j=1}^{t_0} 1/j^\beta)$ for some $t_0 = O(1)$. In this section we will give an explicit upper bound, based on our techniques of estimating sizes and volumes of intervals in $(\alpha,\beta)$-PLG. The lower bound on the size of a dominating set in $\mathcal{G}_{\alpha,\beta}$ given in part (b) of the following lemma was also used by Shen et al.

**Lemma 2.**

*(a) If* $\mathrm{vol}([x\Delta,\Delta]) = \sum_{j=x\Delta}^{\Delta} \left\lfloor \frac{e^\alpha}{j^\beta} \right\rfloor \cdot j < \lfloor e^\alpha \rfloor$, *then* $|[x\Delta,\Delta]|$ *is a lower bound on the size of a dominating set in* $\mathcal{G}_{\alpha,\beta}$.

*(b) If* $\mathrm{vol}([x\Delta,\Delta]) = \sum_{j=x\Delta}^{\Delta} \left\lfloor \frac{e^\alpha}{j^\beta} \right\rfloor \cdot j < \sum_{j=1}^{x\Delta-1} \left\lfloor \frac{e^\alpha}{j^\beta} \right\rfloor$, *then* $|[x\Delta,\Delta]|$ *is a lower bound on the size of a dominating set in* $\mathcal{G}_{\alpha,\beta}$.

*Proof.* Considering (a), let $D$ be a dominating set in $\mathcal{G}_{\alpha,\beta}$, and let $D_1 = D \cap [x\Delta,\Delta]$ and $D_2 = D \backslash D_1$. Suppose $|D_2| < |[x\Delta,\Delta] \backslash D_1|$. Since $\forall v \in D_2, u \in [x\Delta,\Delta]\backslash D_1$ we have $\mathsf{deg}_{\alpha,\beta}(v) < \mathsf{deg}_{\alpha,\beta}(u)$, this implies $\mathrm{vol}(D_2) < \mathrm{vol}([x\Delta,\Delta] \setminus D_1)$ and thus $\mathrm{vol}(D) < \mathrm{vol}([x\Delta,\Delta]) < \lfloor e^\alpha \rfloor$, a contradiction.

Suppose in case (b) that $\mathrm{vol}([x\Delta,\Delta]) < |[1, x\Delta - 1]|$ and that $D, D_1, D_2$ are the same as in the proof of (a). Again we obtain $\mathrm{vol}(D_2) < \mathrm{vol}([x\Delta,\Delta] \setminus D_1)$, which implies $\mathrm{vol}(D) < \mathrm{vol}([x\Delta,\Delta]) < |[1, x\Delta - 1]$. Thus the volume of $D$ is not sufficient to dominate the subset $[1, x\Delta - 1]$, a contradiction. □

In order to demonstrate the power and the limitations of this lower bound, we want to determine the value $\min\{x | \mathrm{vol}([x\Delta,\Delta]) < \lfloor e^\alpha \rfloor\}$.

In the case $\beta > 2$ we consider the following estimates of sizes of intervals and the node degree available in such intervals:

$$\begin{aligned}
\sum_{x\Delta}^{y\Delta} \frac{1}{j^{\beta-1}} &\in \left[\int\limits_{x\Delta}^{y\Delta} \frac{1}{j^{\beta-1}}\,\mathrm{d}j - \left(\frac{1}{(x\Delta)^{\beta-1}} - \frac{1}{(y\Delta)^{\beta-1}}\right), \int\limits_{x\Delta}^{y\Delta} \frac{1}{j^{\beta-1}}\,\mathrm{d}j\right] \\
&= \left[\left[\frac{j^{2-\beta}}{2-\beta}\right]_{x\Delta}^{y\Delta} - \left(\frac{1}{(x\Delta)^{\beta-1}} - \frac{1}{(y\Delta)^{\beta-1}}\right), \left[\frac{j^{2-\beta}}{2-\beta}\right]_{x\Delta}^{y\Delta}\right] \\
&= \left[\frac{y^{2-\beta}-x^{2-\beta}}{2-\beta}\Delta^{2-\beta} - \left(\left(x^{1-\beta}-y^{1-\beta}\right)\Delta^{1-\beta}\right), \frac{y^{2-\beta}-x^{2-\beta}}{2-\beta}\Delta^{2-\beta}\right]
\end{aligned}$$

For the size of the interval $|[x\Delta, y\Delta]| = \sum_{x\Delta}^{y\Delta} \frac{e^\alpha}{j^\beta}$ we get:

$$\begin{aligned}
|[x\Delta, y\Delta]| &\in e^\alpha \left[\frac{\Delta^{1-\beta}}{1-\beta}\left(y^{1-\beta}-x^{1-\beta}\right) - \Delta^{-\beta}\left(\frac{1}{x^\beta}-\frac{1}{y^\beta}\right), \frac{\Delta^{1-\beta}}{1-\beta}\left(y^{1-\beta}-x^{1-\beta}\right)\right] \\
&= e^\alpha \cdot e^{\alpha\frac{1-\beta}{\beta}}\left[\frac{x^{1-\beta}-y^{1-\beta}}{\beta-1} - \frac{1}{\Delta}\left(\frac{1}{x^\beta}-\frac{1}{y^\beta}\right), \frac{x^{1-\beta}-y^{1-\beta}}{\beta-1}\right] \\
&= \left[\Delta\frac{x^{1-\beta}-y^{1-\beta}}{\beta-1} - \left(\frac{1}{x^\beta}-\frac{1}{y^\beta}\right), \Delta\frac{x^{1-\beta}-y^{1-\beta}}{\beta-1}\right]
\end{aligned}$$

We consider the case $x = \frac{2}{\Delta}, y = 1$. We obtain $|[2, \Delta]| = \zeta(\beta)e^\alpha - e^\alpha = (\zeta(\beta)-1)\cdot e^\alpha$ and

$$\sum_{j=2}^{\Delta} \frac{e^\alpha}{j^\beta}\cdot j \in \left[e^\alpha \cdot \left(\frac{\left(\frac{\Delta}{2}\right)^{\beta-2}-1}{\beta-2}\cdot\Delta^{2-\beta} - \left(\left(\left(\frac{\Delta}{2}\right)^{\beta-1}-1\right)\cdot\Delta^{1-\beta}\right)\right), e^\alpha\cdot\frac{\left(\frac{\Delta}{2}\right)^{\beta-2}-1}{\beta-2}\cdot\Delta^{2-\beta}\right]$$

For the interval $[d, \Delta]$ we obtain:

$$\sum_{j=d}^{\Delta}\left\lfloor\frac{e^\alpha}{j^\beta}\right\rfloor\cdot j \leqslant \Delta^2\cdot\frac{\left(\frac{\Delta}{d}\right)^{\beta-2}-1}{\beta-2} = \frac{1}{\beta-2}\cdot\left(\frac{e^{\frac{2\alpha}{\beta}}\cdot e^{\alpha\cdot\frac{\beta-2}{\beta}}}{d^{\beta-2}} - e^{\frac{2\alpha}{\beta}}\right) = (1-o(1))\cdot\frac{e^\alpha}{d^{\beta-2}\cdot(\beta-2)}$$

We obtain the following estimate for the size of the interval $[1, d-1]$:

$$\begin{aligned}
|[1, d-1]| &= \sum_{j=1}^{d}\left\lfloor\frac{e^\alpha}{j^\beta}\right\rfloor \geqslant \sum_{j=1}^{d}\frac{e^\alpha}{j^\beta} - (d-1) \\
&\geqslant e^\alpha\cdot\left(\int\limits_{1}^{d-1} j^{-\beta}\,\mathrm{d}j - \left(1-\frac{1}{(d-1)^\beta}\right)\right) - (d-1) \\
&= e^\alpha\cdot\left(\frac{(d-1)^{1-\beta}-1}{1-\beta} - \left(1-\frac{1}{(d-1)^\beta}\right)\right) - (d-1) \\
&= e^\alpha\cdot\left(\frac{1-\frac{1}{(d-1)^{\beta-1}}}{\beta-1} - 1 + \frac{1}{(d-1)^\beta}\right) - (d-1)
\end{aligned}$$

Hence we want to determine the smallest $d \geqslant 2$ such that

$$\frac{1}{d^{\beta-2}\cdot(\beta-2)} < \frac{(d-1)^\beta - (d-1) - (\beta-1)(d-1)^\beta + \beta - 1}{(\beta-1)(d-1)^\beta}$$

We observe that $1+\frac{1}{2^\beta} > \frac{1}{3^{\beta-2}(\beta-2)}$ for $\beta \geqslant \beta_2 \approx 2.48$, $1+\frac{1}{2^\beta}+\frac{1}{3^\beta} > \frac{1}{4^{\beta-2}(\beta-2)}$ for $\beta \geqslant \beta_3 \approx 2.44$ and $1+\frac{1}{2^\beta}+\frac{1}{3^\beta}+\frac{1}{4^\beta} > \frac{1}{5^{\beta-2}(\beta-2)}$ for $\beta \geqslant \beta_4 \approx 2.40$. This gives the following upper bounds for the approximability of MIN-DS on $(\alpha, \beta)$-PLG for $\beta > 2$.

**Lemma 3.** *For $k \in \{2,3,4\}$ let $\beta_k = \min\left\{\beta \,\middle|\, \sum_{j=1}^{k} \frac{1}{j^\beta} > \frac{1}{k^{\beta-2}(\beta-2)}\right\}$. Then $\beta_2 \approx 2.48$, $\beta_3 \approx 2.44$ and $\beta_3 \approx 2.40$. For $k \in \{2,3,4\}$, for $\beta \geqslant \beta_k$, the* MINIMUM DOMINATING SET *problem in $(\alpha,\beta)$-PLG is hard to approximate within approximation ratio $\left(\zeta(\beta)-\frac{1}{2}\right)\cdot(\beta-2)\cdot(k+1)^{\beta-2}$.*

## 7.1 Improved Analysis

We will now significantly improve the analysis based on the lower bounds from Lemma 2. Instead of just giving upper and lower bounds on the size of an optimum dominating set and a greedy solution separately, we will explicitly relate upper and lower bound to each other.

Let $\mathcal{G}_{\alpha,\beta}$ be an $(\alpha,\beta)$-PLG with $\beta > 2$. Let $W$ be the set of neighbors of degree-1 nodes of degree at least 2 in $\mathcal{G}_{\alpha,\beta}$ and let $M$ be the set of degree-1 nodes in $\mathcal{G}_{\alpha,\beta}$ which are adjacent to another degree-1 node. Let $R = V \setminus (W \cup \{v | \mathsf{deg}_{\alpha,\beta}(v) = 1\})$.

Then there exists some $c = c_\beta > 0$ not depending on $\alpha$ such that $|W| \geqslant c \cdot e^\alpha$. This implies $|R| \leqslant (\zeta(\beta) - c - 1)e^\alpha$.

**Lemma 4.** *If $\mathcal{G}_{\alpha,\beta}$ is a connected $(\alpha,\beta)$-PLG with $\beta > 2$ and $W$ and $R$ are defined as above, then there exists an optimum dominating set $OPT$ in $\mathcal{G}_{\alpha,\beta}$ with $OPT = OPT_R \cup W \cup M'$, where $OPT_R$ is an optimum dominating set for the induced subgraph $\mathcal{G}_{\alpha,\beta}[R]$ on $R$ and $M' \subset M$ is of cardinality $|M'| = \frac{|M|}{2}$.*

The maximum degree in $\mathcal{G}_{\alpha,\beta}[R]$ is at most $\Delta$. We consider the dominating set $D = W \cup D_{Gr} \cup M'$ where $D_{Gr}$ is a dominating set for $\mathcal{G}_{\alpha,\beta}[R]$ constructed by the Greedy Algorithm and $M' \subset M$ is a subset of size $\frac{|M|}{2}$ dominating $M$. The approximation ratio is at most

$$\frac{\ln(\Delta+1)\cdot|\mathsf{OPT}_R| + |W| + \frac{|M|}{2}}{|\mathsf{OPT}_R| + |W| + \frac{|M|}{2}} \leqslant \frac{\frac{\alpha}{\beta}\cdot|\mathsf{OPT}_R| + |W| + \frac{|M|}{2}}{|\mathsf{OPT}_R| + |W| + \frac{|M|}{2}}$$

We can further improve this bound as follows. Since $R = V \setminus (W \cup V_1)$ and $|\mathsf{OPT}_R| \leqslant |R|$, the approximation ratio is at most

$$\max\left\{\frac{r\cdot|\mathsf{OPT}_R| + |W| + \frac{|M|}{2}}{|\mathsf{OPT}_R| + |W| + \frac{|M|}{2}} \,\middle|\, \begin{array}{l}|\mathsf{OPT}_R| \leqslant |R| \\ r = \min\left\{\frac{\alpha}{\beta}, \frac{|R|}{|\mathsf{OPT}_R|}\right\}\end{array}\right\}$$

**Case 1:** $\left(\mathbf{r} = \frac{\alpha}{\beta}\right)$ This means that $\frac{\alpha}{\beta} \leqslant \frac{|R|}{|\mathsf{OPT}_R|}$, i.e. $|\mathsf{OPT}_R| \leqslant \frac{\beta}{\alpha}\cdot|R|$. The upper bound for the approximation ratio is monotone increasing in $|\mathsf{OPT}_R|$, hence it is bounded by

$$\frac{\frac{\alpha}{\beta}\cdot\frac{\beta}{\alpha}\cdot|R| + |W| + \frac{|M|}{2}}{\frac{\beta}{\alpha}\cdot|R| + |W| + \frac{|M|}{2}} = \frac{|R| + |W| + \frac{|M|}{2}}{\frac{\beta}{\alpha}\cdot|R| + |W| + \frac{|M|}{2}}$$

**Case 2:** $\left(\mathbf{r} = \frac{|\mathbf{R}|}{|\mathsf{OPT}_\mathbf{R}|} < \frac{\alpha}{\beta}\right)$ Then we have $|\mathsf{OPT}_R| > \frac{\beta\cdot|R|}{\alpha}$, and we obtain

$$\frac{r\cdot|\mathsf{OPT}_R| + |W| + \frac{|M|}{2}}{|\mathsf{OPT}_R| + |W| + \frac{|M|}{2}} = \frac{|R| + |W| + \frac{|M|}{2}}{|\mathsf{OPT}_R| + |W| + \frac{|M|}{2}} \leqslant \frac{|R| + |W| + \frac{|M|}{2}}{\frac{\beta}{\alpha}\cdot|R| + |W| + \frac{|M|}{2}}$$

Now we need to construct an upper bound for the term $\frac{|R|+|W|+\frac{|M|}{2}}{\frac{\beta}{\alpha}\cdot|R|+|W|+\frac{|M|}{2}}$. Recall that the volume of a set of nodes $U \subseteq V$ is defined as $\mathrm{vol}(U) = \sum_{u\in U} \mathsf{deg}_{\alpha,\beta}(u)$. We consider two cases.

**Case I:** $(\zeta(\beta-1)-1<1)$ In this case, the volume of nodes of degree at least two does not suffice to dominate all the degree-1 nodes. Hence in this case, $M \neq \emptyset$. We obtain the following lower bound for the cardinality of $M$: $|M| \geqslant e^\alpha - (\zeta(\beta-1)-1)e^\alpha = (2-\zeta(\beta-1))e^\alpha$. Nevertheless we will use the upper bound $|R| \leqslant (\zeta(\beta)-1)e^\alpha$. Since the term $\frac{|R|+|W|+\frac{|M|}{2}}{\frac{\beta}{\alpha}\cdot|R|+|W|+\frac{|M|}{2}}$ is monotone increasing in $|R|$, we obtain

$$\rho(\beta) = \frac{|R|+|W|+\frac{|M|}{2}}{\frac{\beta}{\alpha}\cdot|R|+|W|+\frac{|M|}{2}} \leqslant \frac{(\zeta(\beta)-1)e^\alpha + \frac{(2-\zeta(\beta-1))e^\alpha}{2}}{\frac{\beta}{\alpha}\cdot\zeta(\beta)-1)e^\alpha + \frac{(2-\zeta(\beta-1))e^\alpha}{2}} = \frac{\zeta(\beta)-\frac{\zeta(\beta-1)}{2}}{1-\frac{\zeta(\beta-1)}{2}}$$

In Figure 3 we plot the above approximation ratio in comparison to the ratio $\frac{\zeta(\beta)-\frac{1}{2}}{\zeta(\beta)-1}$ of Shen et al. [She+12] for $\beta \geqslant 2.75$.

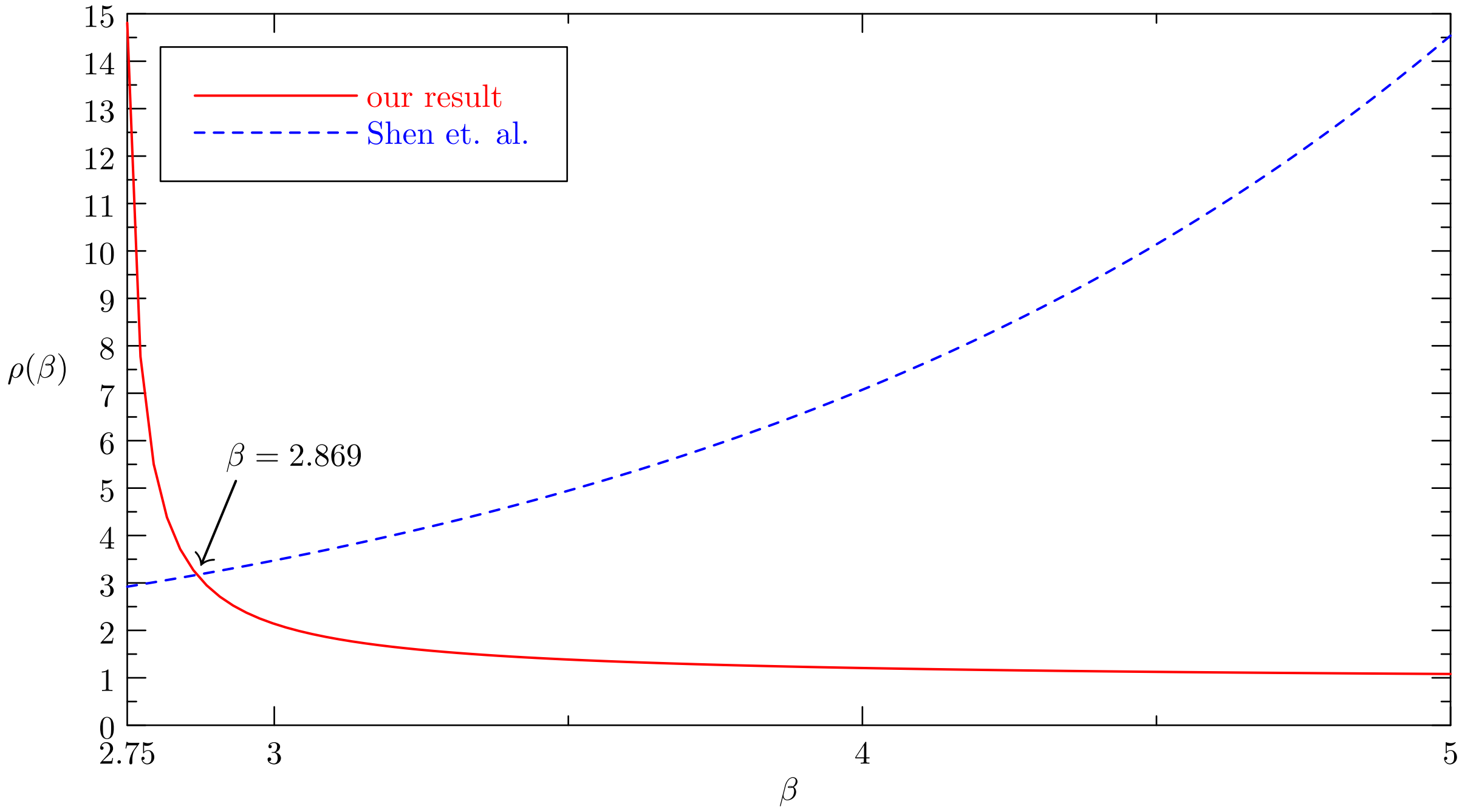

**Figure 3:** Plot of the approximation ratios $\frac{\zeta(\beta)-\frac{\zeta(\beta-1)}{2}}{1-\frac{\zeta(\beta-1)}{2}}$ (our result) in comparison to $\frac{\zeta(\beta)-\frac{1}{2}}{\zeta(\beta)-1}$ (Shen et al.) for $\beta \geqslant 2.75$.

**Case II:** $(\zeta(\beta-1)-1 \geqslant 1)$ In this case, the volume of the nodes of degree at least 2 suffices to dominate the degree-1 nodes. Now we construct a lower bound for $|W|$ as follows:

$$\begin{aligned}|W| &\geqslant \min\left\{|[d,\Delta]| \mid \mathrm{vol}([d,\Delta]) > e^\alpha\right\} \\ &= \min\left\{\left(\zeta(\beta) - \sum_{j=1}^{d-1}\frac{1}{j^\beta}\right)e^\alpha \,\middle|\, \left(\zeta(\beta-1) - \sum_{j=1}^{d-1}\frac{1}{j^{\beta-1}}\right)e^\alpha > e^\alpha\right\}\end{aligned}$$

Hence in this case the approximation ratio is bounded by

$$\frac{\zeta(\beta)-1}{\frac{\beta}{\alpha}\cdot|[1,d-1]|+|[d,\Delta]|} = \frac{\zeta(\beta)-1}{\zeta(\beta)-\sum_{j=1}^{d-1}\frac{1}{j^\beta}}$$

where $d = \min\{d' | \mathrm{vol}([d',\Delta]) > e^\alpha\}$.

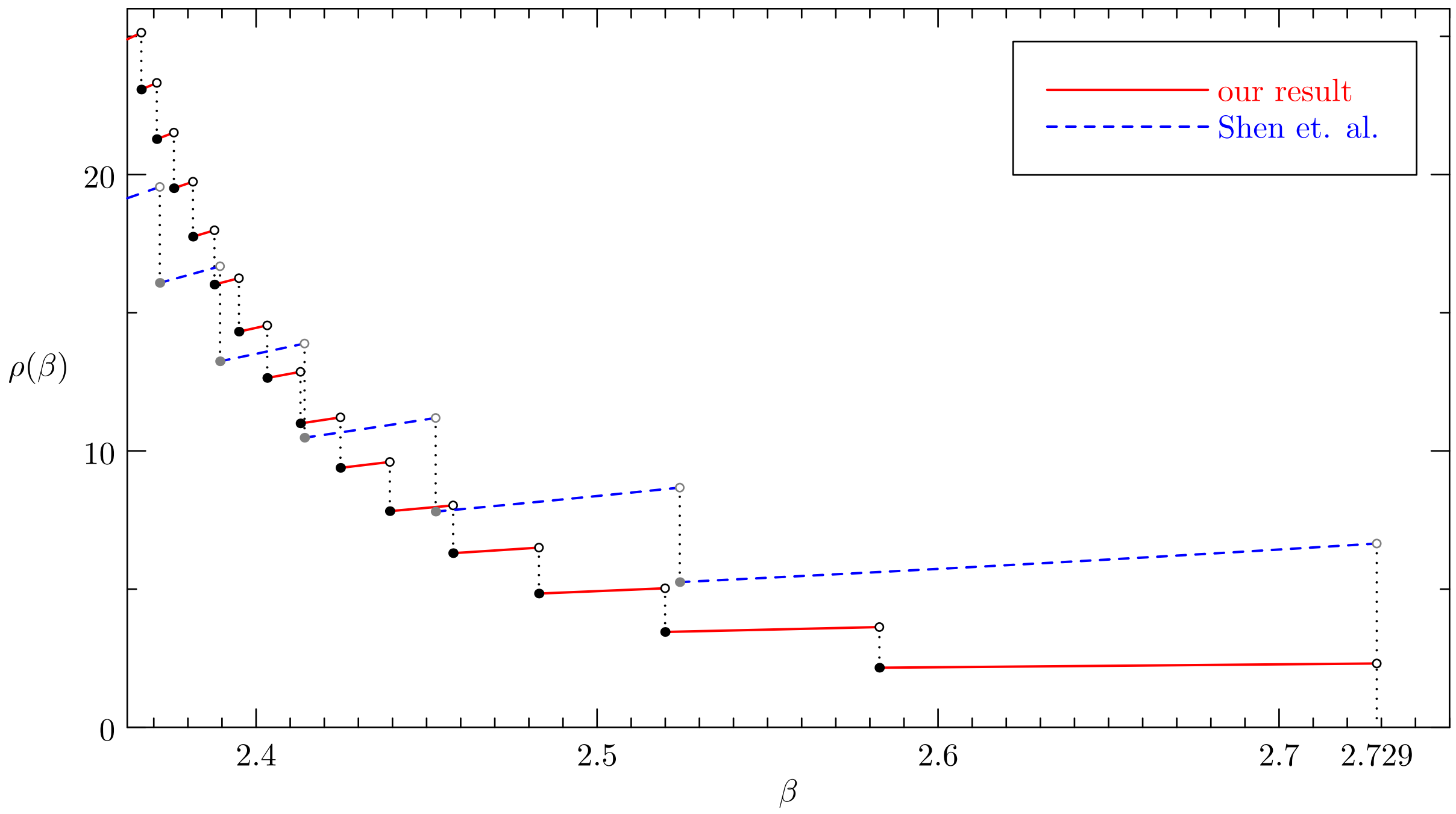

**Figure 4:** Comparison of our approximation ratio to the previous approximation ratio of Shen et al.

## 8 The Functional Case $\beta_f = 2 + \frac{1}{f(n)}$

We consider now the case when the parameter $\beta$ is a function of the size $n$ of the power law graph, converging to 2 from above. In the preceding sections we have shown that for $\beta \leqslant 2$, there is a logarithmic lower bound for the approximability of the Minimum Dominating Set problem in $(\alpha, \beta)$-PLG. On the other hand, for $\beta > 2$ the problem is in APX (cf. Shen et al. [She+12] and the previous section). Thus we may now have a closer look at this phase transition at $\beta = 2$. Similar as in our previous paper we consider the case when $\beta$ is a function of the size $n$ of the power law graph such that this function converges to 2 from above. Surprisingly we will obtain a very tight phase transition of the computational complexity of the problem, depending on the convergence rate of the function. Let us first give a precise description of the model.

**Definition 1.** **($(\alpha, \beta_f)$-PLG for $\beta_f = 2 + \frac{1}{f(n)}$)**
Let $f \colon \mathbb{N} \to \mathbb{N}$ be a monotone increasing unbounded function. For $\beta_f = 2 + \frac{1}{f(n)}$, an $(\alpha, \beta_f)$-PLG is an undirected multigraph $\mathcal{G}_{\alpha,\beta_f}$ with $n$ nodes and maximum degree $\Delta_f = \left\lfloor e^{\alpha/\beta_f} \right\rfloor$ such that for $j = 1, \ldots, \Delta_f = \left\lfloor e^{\alpha/\beta_f} \right\rfloor$, the number of nodes of degree $j$ in $\mathcal{G}_{\alpha,\beta_f}$ equals $\left\lfloor \frac{e^\alpha}{j^{\beta_f}} \right\rfloor$.

Especially this means that $\sum_{j=1}^{\Delta_f} \left\lfloor \frac{e^\alpha}{j^{2+1/f(n)}} \right\rfloor = n$.

In order to study the computational complexity of the Minimum Dominating Set problem in $(\alpha, \beta_f)$-power law graphs, we start again by giving sufficiently precise estimates of sizes of intervals.

**Convergence of terms $\mathbf{j^{-\beta_f}}$.** First we give an additive bound for the terms $j^{-\beta_f}$. $\frac{1}{j^{\beta_f}} = \frac{1}{j^{2+\frac{1}{f(n)}}} \in \left[\frac{1}{j^2} - \tau(n), \frac{1}{j^2}\right]$, where

$$\tau(n) = \max\left\{ \frac{1}{j^2} - \frac{1}{j^{2-\frac{1}{f(n)}}} \,\middle|\, j = 1, \ldots, \Delta_f \right\} = \max\left\{ = \frac{j^{\frac{1}{f(n)}} - 1}{j^{2+\frac{1}{f(n)}}} \,\middle|\, j = 1, \ldots, \Delta_f \right\}$$

We consider the function $x \mapsto h(x) := \frac{x^{\frac{1}{f(n)}} - 1}{x^{2+\frac{1}{f(n)}}} = x^{-2} - x^{-2-\frac{1}{f(n)}}$. Its derivative is $\frac{d}{dx} h(x) = \frac{d}{dx} \frac{x^{\frac{1}{f(n)}} - 1}{x^{2+\frac{1}{f(n)}}} = -2x^{-3} + \left(2 + \frac{1}{f(n)}\right) x^{-3-\frac{1}{f(n)}}$. The condition $h(x) < 0$ is equivalent to $1 + \frac{1}{2f(n)} < x^{\frac{1}{f(n)}}$. We observe that the derivative attains its maximum at $x = 2$. We have

$$h'(2) < 0 \iff \left(1 + \frac{1}{2f(n)}\right)^{f(n)} < 2$$

We observe that $\lim_{n\to\infty} \left(1 + \frac{1}{2f(n)}\right)^{f(n)} = e^{1/2} < 2$. Thus we obtain $\tau(n) = \frac{2^{1/f(n)} - 1}{2^{2+1/f(n)}}$.

Now we give a multiplicative bound as follows. We have

$$\frac{1}{j^{\beta_f}} = \frac{1}{j^2} \cdot j^{2-\beta_f} = \frac{1}{j^2} \cdot \frac{1}{j^{1/f(n)}} \in \left[\frac{1}{n^{1/f(n)}} \cdot \frac{1}{j^2}, \frac{1}{j^2}\right]$$

**Sizes of Intervals.** For $\beta = 2$, our technique based on integration yields the following estimate of sizes of intervals:

$$\begin{aligned}
\sum_{j=x\Delta}^{y\Delta} \frac{1}{j^2} &\in \left[\int_{x\Delta}^{y\Delta} j^{-2}\,\mathrm{d}j,\ \int_{x\Delta}^{y\Delta} j^{-2}\,\mathrm{d}j + \frac{1}{(x\Delta)^2} - \frac{1}{(y\Delta)^2}\right] \\
&= \left[\frac{1}{x\Delta} - \frac{1}{y\Delta}, \frac{1}{x\Delta} - \frac{1}{y\Delta} + \frac{1}{(x\Delta)^2} - \frac{1}{(y\Delta)^2}\right] \\
|[x\Delta, y\Delta]| &\in \left[e^{\alpha/2} \cdot \left(\frac{1}{x} - \frac{1}{y}\right),\ e^{\alpha/2} \cdot \left(\frac{1}{x} - \frac{1}{y}\right) + \frac{1}{x^2} - \frac{1}{y^2}\right]
\end{aligned}$$

We combine this with the multiplicative bound and obtain the following estimate of the size of intervals in the case $\beta_f = 2 + \frac{1}{f(n)}$.

$$\begin{aligned}
|[x\Delta_f, y\Delta_f]| &= \sum_{j=x\Delta_f}^{y\Delta_f} \left\lfloor \frac{e^\alpha}{j^{\beta_f}} \right\rfloor \\
&\in \left[ e^{\alpha \cdot \frac{1+\frac{1}{f(n)}}{2+\frac{1}{f(n)}}} \cdot \left(\frac{1}{x} - \frac{1}{y}\right) - (y-x)\Delta_f,\ e^{\alpha \cdot \frac{1+\frac{1}{f(n)}}{2+\frac{1}{f(n)}}} \cdot \left(\frac{1}{x} - \frac{1}{y}\right) + e^{\alpha \cdot \left(1 - \frac{1}{1+\frac{1}{2f(n)}}\right)} \cdot \left(\frac{1}{x^2} - \frac{1}{y^2}\right) \right] \\
&= \left[ e^{\alpha \cdot \frac{f(n)+1}{2f(n)+1}} \cdot \left(\frac{1}{x} - \frac{1}{y}\right) - (y-x)\Delta_f,\ e^{\alpha \cdot \frac{f(n)+1}{2f(n)+1}} \cdot \left(\frac{1}{x} - \frac{1}{y}\right) + e^{\alpha \cdot \frac{1}{2f(n)+1}} \cdot \left(\frac{1}{x^2} - \frac{1}{y^2}\right) \right]
\end{aligned}$$

Especially we obtain the following estimate of the size of $\mathcal{G}_{\alpha,\beta_f}$:

$$\begin{aligned}
|[1, \Delta_f]| &\in \left[e^\alpha - e^{\alpha\frac{f(n)+1}{2f(n)+1}} - e^{\alpha\frac{f(n)}{2f(n)+1}} + 1,\ e^\alpha - e^{\alpha\frac{f(n)+1}{2f(n)+1}} + e^{\alpha\frac{1}{2f(n)+1}} \cdot e^{2\alpha\frac{f(n)}{2f(n)+1}} - e^{\alpha\frac{1}{2f(n)+1}}\right] \\
&= [(1 - o(1))e^\alpha,\ (2 - o(1))e^\alpha]
\end{aligned}$$

This estimate can be refined as follows:

$$\sum_{j=1}^{\Delta_f} \left\lfloor \frac{e^\alpha}{j^{\beta_f}} \right\rfloor \in \left[ \sum_{j=1}^{\Delta_f} \frac{e^\alpha}{j^{\beta_f}} - \Delta_f, \sum_{j=1}^{\Delta_f} \frac{e^\alpha}{j^{\beta_f}} \right]$$

$$\subseteq \left[ \frac{1}{n^{1/f(n)}} \cdot \sum_{j=1}^{\Delta_f} \frac{e^\alpha}{j^2} - \Delta_f, \sum_{j=1}^{\Delta_f} \frac{e^\alpha}{j^2} \right] \subseteq [(1 - o(1))\zeta(2)e^\alpha, \zeta(2)e^\alpha]$$

where the last inclusion holds for $f(n) = \omega(\log(\alpha))$. The volume can be estimated as follows:

$$\begin{aligned}
\mathrm{vol}([x\Delta_f, y\Delta_f]) &= \sum_{x\Delta_f}^{y\Delta_f} \left\lfloor \frac{e^\alpha}{j^{\beta_f}} \right\rfloor \cdot j \\
&\in \left[ \sum_{x\Delta_f}^{y\Delta_f} \frac{e^\alpha}{j^{\beta_f - 1}} - (x\Delta_f + (x\Delta_f + 1) + \ldots + y\Delta_f), \sum_{x\Delta_f}^{y\Delta_f} \frac{e^\alpha}{j^{\beta_f - 1}} \right] \\
&= \left[ \sum_{x\Delta_f}^{y\Delta_f} \frac{e^\alpha}{j^{\beta_f - 1}} - \frac{(y^2 - x^2)\Delta_f^2 + (x + y)\Delta_f}{2}, \sum_{x\Delta_f}^{y\Delta_f} \frac{e^\alpha}{j^{\beta_f - 1}} \right]
\end{aligned}$$

Since $j^{\beta_f - 1} = j^{1 + \frac{1}{f(n)}}, j = x\Delta_f, y\Delta_f$, we use Lemma 23 from our previous paper and obtain that the volume $\mathrm{vol}([x\Delta_f, y\Delta_f])$ is within the interval

$$\left[ \frac{e^\alpha \cdot (\ln(y) - \ln(x))}{n^{\frac{1}{f(n)}}} - \frac{(y^2 - x^2)\Delta_f^2 + (x + y)\Delta_f}{2}, e^\alpha \cdot (\ln(y) - \ln(x)) + e^\alpha \cdot \left( \frac{1}{x\Delta_f} - \frac{1}{y\Delta_f} \right) \right]$$

We are now well prepared to compute the parameters $\alpha, d, x, y$ of our embedding $G_{U,\mathcal{S}} \mapsto \mathcal{G}_{\alpha,\beta_f}$ for the functional case $\beta_f = 2 + \frac{1}{f(n)}$, $f(n) = \omega(\log(n))$. Recall that we want to choose $\alpha, d, x, y$ such as to meet the following requirements:

(1) $\left| \left[ N^{\frac{a}{d}}, N^{\frac{b}{d}} \right] \right| \geqslant N$

(2) $|[x\Delta_f, y\Delta_f]| = o\left( N^{\frac{d-1}{d}} \right)$
(where $N^{\frac{d-1}{d}}$ is a lower bound for the size of an opt. dom.set in $G_{U,\mathcal{S}}$)

(3) $\sum_{j=x\Delta_f}^{y\Delta_f} \left\lfloor \frac{e^\alpha}{j^{\beta_f}} \right\rfloor \cdot j = \mathrm{vol}\left( [x\Delta_f, y\Delta_f] \right) \geqslant \zeta(2) \cdot e^\alpha$

We want to get an estimate for $\left| \left[ N^{\frac{a}{d}}, N^{\frac{b}{d}} \right] \right|$. Note that $e^{\alpha \cdot \frac{1}{2f(n)-1}} \cdot \Delta_f^2 = e^{\alpha \cdot \frac{f(n)+1}{2f(n)+1}} \cdot \Delta_f = e^\alpha$. Thus our estimate of interval sizes yields

$$\left| \left[ N^{\frac{a}{d}}, N^{\frac{b}{d}} \right] \right| \in \left[ e^\alpha \left( \frac{1}{N^{\frac{a}{d}}} - \frac{1}{N^{\frac{b}{d}}} \right) - \left( N^{\frac{b}{d}} - N^{\frac{a}{d}} \right), e^\alpha \left( \frac{1}{N^{\frac{a}{d}}} - \frac{1}{N^{\frac{b}{d}}} \right) + e^\alpha \left( \frac{1}{N^{\frac{2a}{d}}} - \frac{1}{N^{\frac{2b}{d}}} \right) \right]$$

In order to satisfy (1), for a given $d$ we want to choose $\alpha$ such that

$$\left| \left[ N^{\frac{a}{d}} \right] \right| \geqslant e^\alpha \left( \frac{1}{N^{\frac{a}{d}}} - \frac{1}{N^{\frac{b}{d}}} \right) - \left( N^{\frac{b}{d}} - N^{\frac{a}{d}} \right) \Longleftrightarrow e^\alpha \left( N^{\frac{b-a}{d}} - 1 \right) - \left( 1 - N^{\frac{a-b}{d}} \right) \geqslant N^{1 + \frac{b}{d}}$$

Hence we choose

$$e^\alpha \approx N^{1 + \frac{a}{d}} \Longleftrightarrow \alpha \approx \left( 1 + \frac{a}{d} \right) \cdot \ln(N)$$

If we now choose $d > \frac{(b+1)\beta_f}{\beta_f - 1}$, then the constraint (2) holds, and for $y = 1$ and $x > 0$ such that $x\Delta_f > N^{b/d}$, constraint (3) holds as well. Thus we obtain asymptotically the same approximation hardness result as for the case $\beta = 2$.

**The Case $\mathbf{f(n) = o(\log(n))}$.** Let us now consider the case when $f(n)$ is a "*slowly growing*" function, namely $f(n) = o(\log(n))$. In that case, $n^{\frac{1}{f(n)}} \longrightarrow \infty$ as $n \longrightarrow \infty$. For $x\Delta_f \leqslant j \leqslant y\Delta_f$ we obtain

$$\frac{1}{j^{1+\frac{1}{f(n)}}} = \frac{1}{j} \cdot \frac{1}{j^{\frac{1}{f(n)}}} \leqslant \frac{1}{j} \cdot \frac{1}{(x\Delta_f)^{\frac{1}{f(n)}}} = \frac{1}{j} \cdot \frac{1}{x^{\frac{1}{f(n)}}} \cdot \frac{1}{e^{\alpha \cdot \frac{1}{2f(n)+1}}}$$

and therefore

$$\mathrm{vol}([x\Delta_f, \Delta_f]) \leqslant e^{\alpha} \cdot \ln\left(\frac{1}{x}\right) \cdot \frac{1}{x^{\frac{1}{f(n)}}} \cdot \frac{1}{e^{\alpha \cdot \frac{1}{2f(n)+1}}}$$

which yields the requirement $\ln\left(\frac{1}{x}\right)/x^{\frac{1}{f(n)}} \;\geqslant\; c \cdot e^{\alpha \cdot \frac{1}{2f(n)+1}}$. This is equivalent to

$$\ln\ln\left(\frac{1}{x}\right) + \frac{1}{f(n)} \cdot \ln\left(\frac{1}{x}\right) \;\geqslant\; \ln(c) + \frac{\alpha}{2f(n)+1}$$

which means the following: In order to dominate the remaining vertices of the graph with vertices from $[x\Delta_f, \Delta_f]$, we have to choose (roughly) $\ln\left(\frac{1}{x}\right) \geqslant \frac{\alpha}{2}$, i.e. $\frac{1}{x} \geqslant e^{\frac{\alpha}{2}}$. This gives the following lower bound for the size of that interval:

$$|[x\Delta_f, \Delta_f]| \;\geqslant\; e^{\alpha \cdot \frac{f(n)+1}{2f(n)+1}} \cdot \left(e^{\frac{\alpha}{2}} - 1\right) - \left(1 - \frac{1}{e^{\frac{\alpha}{2}}}\right) \cdot e^{\frac{\alpha}{2+\frac{1}{f(n)}}} \;\geqslant\; (1 - o(1))e^{\frac{\alpha}{2} \cdot \left(1 + \frac{f(n)+1}{f(n)+1/2}\right)}$$

This lower bound for the size of $[x\Delta_f, \Delta_f]$ converges to $e^{\alpha}$ as $n \to \infty$, which means there exists some $c > 0$ such that $|[x\Delta_f, \Delta_f]| \geqslant c \cdot |\mathcal{G}_{\alpha,\beta_f}|$ in order to be a dominating set. Hence each dominating set in $\mathcal{G}_{\alpha,\beta_f}$ is of cardinality at least $c \cdot |\mathcal{G}_{\alpha,\beta_f}|$. Thus we obtain the following result.

**Theorem 5.** *For $\beta_f = 2 + \frac{1}{f(n)}$ with $f(n) = o(\log(n))$, the* Minimum Dominating Set *problem on $(\alpha, \beta_f)$-PLG is in APX.*

# 9 Further Research

The possible further improvements in both lower and upper approximation bounds are important open problems in the area. As some of our lower approximation bounds are asymptotically optimal, the challenging questions remain on possible improvements in upper approximation bounds for the cases which are without optimal bounds. Another interesting problem concerns the approximability of PLG optimization problems in a random or quasi-random model.

# References

[ACL00] W. Aiello, F. Chung, and L. Lu. "A random graph model for massive graphs". *Proceedings of the 32nd annual ACM Symposium on Theory of Computing.* STOC '00. ACM Press, 2000, pp. 171–180. DOI: 10.1145/335305.335326.

[ACL01] W. Aiello, F. Chung, and L. Lu. "A random graph model for power law graphs". *Experimental Mathematics* 10.1 (2001), pp. 53–66.

[BM84] R. Bar-Yehuda and S. Moran. "On approximation problems related to the independent set and vertex cover problems". *Discrete Applied Mathematics* 9.1 (09/1984), pp. 1–10. DOI: 10.1016/0166-218X(84)90086-6.

[Bro+00] A. Broder, R. Kumar, F. Maghoul, P. Raghavan, S. Rajagopalan, R. Stata, A. Tomkins, and J. Wiener. "Graph structure in the Web". *Computer Networks* 33.1-6 (06/2000), pp. 309–320. DOI: 10.1016/S1389-1286(00)00083-9.

[ER60] P. Erd??s and A. Rényi. "On the evolution of random graphs". *Publications of the Mathematical Institute of the Hungarian Academy of Sciences* 5 (1960), pp. 17–61.

[Eub+04a] S. Eubank, H. Guclu, V. S. A. Kumar, M. V. Marathe, A. Srinivasan, Z. Toroczkai, and N. Wang. "Modelling disease outbreaks in realistic urban social networks". *Nature* 429.6988 (05/2004), pp. 180–184. DOI: 10.1038/nature02541.

[Eub+04b] S. Eubank, V. S. A. Kumar, M. V. Marathe, A. Srinivasan, and N. Wang. "Structural and algorithmic aspects of massive social networks". *Proceedings of the 15th annual ACM-SIAM Symposium on Discrete Algorithms*. SODA '04. SIAM, 2004, pp. 718–727. DOI: 10.1145/982792.982902.

[Fei98] U. Feige. "A threshold of ln n for approximating set cover". *Journal of the ACM* 45.4 (07/1998), pp. 634–652. DOI: 10.1145/285055.285059.

[FFF99] M. Faloutsos, P. Faloutsos, and C. Faloutsos. "On power-law relationships of the Internet topology". *ACM SIGCOMM Computer Communication Review* 29.4 (10/1999), pp. 251–262. DOI: 10.1145/316194.316229.

[FPP08] A. Ferrante, G. Pandurangan, and K. Park. "On the hardness of optimization in power-law graphs". *Theoretical Computer Science* 393.1-3 (03/2008), pp. 220–230. DOI: 10.1016/j.tcs.2007.12.007.

[GHK12] M. Gast, M. Hauptmann, and M. Karpinski. "Improved Approximation Lower Bounds for Vertex Cover on Power Law Graphs and Some Generalizations". *Arxiv preprint arXiv:1210.2698* (10/2012), pp. 1–26. arXiv:1210.2698.

[Gue+02] N. Guelzim, S. Bottani, P. Bourgine, and F. Képès. "Topological and causal structure of the yeast transcriptional regulatory network." *Nature Genetics* 31.1 (05/2002), pp. 60–63. DOI: 10.1038/ng873.

[JAB01] M. Jovanović, F. S. Annexstein, and K. A. Berman. "Modeling peer-to-peer network topologies through "small-world" models and power laws". *IX Telecommunications Forum, TELFOR*. 2001, pp. 1–4.

[Kan92] V. Kann. "On the approximability of NP-complete optimization problems". PhD thesis. Royal Institute of Technology, Stockholm, 1992.

[KL01] J. M. Kleinberg and S. Lawrence. "The structure of the Web." *Science* 294.5548 (11/2001), pp. 1849–1850. DOI: 10.1126/science.1067014.

[Kle+99] J. M. Kleinberg, R. Kumar, P. Raghavan, S. Rajagopalan, and A. S. Tomkins. "The Web as a graph: measurements, models and methods". *Proceedings of the 5th annual international Conference on Computing and Combinatorics*. COCOON '99. Springer-Verlag, 1999, pp. 1–17. DOI: 10.1007/3-540-48686-0_1.

[Kum+00] R. Kumar, P. Raghavan, S. Rajagopalan, D. Sivakumar, A. S. Tomkins, and E. Upfal. "Stochastic models for the Web graph". *Proceedings of the 41st annual Symposium on Foundations of Computer Science*. FOCS '00. IEEE Comput. Soc. Press, 2000, pp. 57–65. DOI: 10.1109/SFCS.2000.892065.

[PM81] A. Paz and S. Moran. "Non deterministic polynomial optimization problems and their approximations". *Theoretical Computer Science* 15.3 (01/1981), pp. 251–277. DOI: 10.1016/0304-3975(81)90081-5.

[PV01] R. Pastor-Satorras and A. Vespignani. “Epidemic Spreading in Scale-Free Networks”. *Physical Review Letters* 86.14 (04/2001), pp. 3200–3203. DOI: 10.1103/PhysRevLett.86.3200.

[RS97] R. Raz and S. Safra. “A sub-constant error-probability low-degree test, and a sub-constant error-probability PCP characterization of NP”. *Proceedings of the 29th annual ACM Symposium on Theory of Computing*. STOC ’97. ACM Press, 05/1997, pp. 475–484. DOI: 10.1145/258533.258641.

[Ses+08] M. Seshadri, S. Machiraju, A. Sridharan, J. Bolot, C. Faloutsos, and J. Leskovec. “Mobile call graphs”. *Proceeding of the 14th ACM SIGKDD international Conference on Knowledge Discovery and Data Mining*. KDD ’08. ACM Press, 08/2008, pp. 596–604. DOI: 10.1145/1401890.1401963.

[She+12] Y. Shen, D. T. Nguyen, Y. Xuan, and M. T. Thai. “New techniques for approximating optimal substructure problems in power-law graphs”. *Theoretical Computer Science* 447 (08/2012), pp. 107–119. DOI: 10.1016/j.tcs.2011.10.023.

[Sig+03] G. Siganos, M. Faloutsos, P. Faloutsos, and C. Faloutsos. “Power laws and the AS-level Internet topology”. *IEEE/ACM Transactions on Networking* 11.4 (08/2003), pp. 514–524. DOI: 10.1109/TNET.2003.815300.

[SNT10] Y. Shen, D. T. Nguyen, and M. T. Thai. “On the hardness and inapproximability of optimization problems on power law graphs”. *Proceedings of the 4th international Conference on Combinatorial Optimization and Applications*. COCOA ’10. Springer-Verlag, 2010, pp. 197–211. DOI: 10.1007/978-3-642-17458-2_17.

[SSZ02] I. Stojmenovic, M. Seddigh, and J. Zunic. “Dominating sets and neighbor elimination-based broadcasting algorithms in wireless networks”. *IEEE Transactions on Parallel and Distributed Systems* 13.1 (2002), pp. 14–25. DOI: 10.1109/71.980024.